\documentclass[notitlepage, a4paper, 10pt]{article}
\usepackage{geometry}

\usepackage{caption}
\usepackage{subcaption}

\usepackage{amsmath}
\usepackage{amssymb}
\usepackage{amsthm}
\usepackage{mathtools}
\usepackage{stmaryrd}
\usepackage{graphicx}
\usepackage{centernot}
\usepackage{gensymb}
\usepackage{marginnote}
\usepackage{wasysym}
\usepackage{physics}
\usepackage{empheq}
\usepackage{tikz}
\usepackage{qcircuit}
\usepackage{color, soul}
\usepackage{xcolor}
\usepackage[ruled, vlined, linesnumbered]{algorithm2e}
\usepackage{hyperref}

\usetikzlibrary{positioning}
\usetikzlibrary{calc}
\usetikzlibrary{matrix}
\usetikzlibrary{decorations.pathreplacing,calligraphy}

\usepackage{tikzit}
\tikzstyle{dot}=[fill=black, draw=black, shape=circle, minimum size=5pt, inner sep=0pt]
\tikzstyle{dot1}=[fill=black, draw=black, shape=circle, minimum size=3pt, inner sep=0pt]
\tikzstyle{block}=[fill=white, draw=black, shape=rectangle, minimum size=20pt, thick]
\tikzstyle{scalar}=[fill=white, draw=black, shape=diamond, minimum size=20pt, thick]
\tikzstyle{tri1}=[fill=white, draw=black, shape=regular polygon, regular polygon sides=3, tikzit shape=rectangle, minimum size=20pt, thick, inner sep=0pt]
\tikzstyle{tri2}=[fill=white, draw=black, shape=regular polygon, regular polygon sides=3, tikzit shape=rectangle, minimum size=20pt, shape border rotate=180, thick, inner sep=0pt]

\tikzstyle{wire}=[-, thick]
\tikzstyle{wirett}=[-, ultra thick]
\tikzstyle{ldashed}=[-, dashed, thick, gray]

\newcounter{mynote}

\newcommand{\mynotenonum}[1]{
	\marginnote{\tiny#1}
}

\newtheorem{theorem}{Proposition}[section]

\newtheorem{lemma}[theorem]{Lemma}
\newtheorem{definition}{Definition}
\theoremstyle{remark}
\newtheorem{remark}{Remark}[section]
\newtheorem*{notation}{Notation}

\newcommand{\symmg}[1]{\mathcal{S}_{#1}}			
\newcommand{\isep}{\mathrel{{.}\,{.}}\nobreak}		
\newcommand{\bvec}[2]{\ensuremath{\mathbf{e}_{#1}^{(#2)}}}	
\newcommand{\twodots}{\mathinner {\ldotp \ldotp}}
\newcommand{\intset}[2]{\ensuremath{\left[#1\twodots{} #2\right]}}
\newcommand{\chridx}{\ensuremath{\chi_e}}			
\newcommand{\gmaxdeg}{\ensuremath{\Delta}}			
\newcommand{\projop}[1]{\ensuremath{\mathfrak{P}_{#1}}}		

\DeclareMathOperator{\vecop}{\mathrm{vec}_r}		
\DeclareMathOperator{\traceop}{Tr}					
\DeclareMathOperator{\cardf}{\mathbf{card}}			
\newcommand{\idenm}[1]{\ensuremath{\mathbb{I}_{#1}}}		
\newcommand{\onesv}[1]{\ensuremath{\mathbf{1}_{#1}}}		

\newcommand{\vectheta}{\ensuremath{\boldsymbol{\theta}}}	
\newcommand{\myendofremark}{\hfill$\triangle$}

\newcommand{\prooflater}{Proof in Appendix \ref{section:proofs}.}

\newcommand{\timestepsset}{\ensuremath{\intset{0}{T-1}}}
\newcommand{\timeent}[2][t]{\ensuremath{{#2}^{(#1)}}}	

\newcommand{\myrom}[1]{\uppercase\expandafter{\romannumeral #1\relax}}

\author{Nicola Mariella\thanks{IBM Quantum, IBM Research Europe - Dublin}\and Sergiy Zhuk\footnotemark[1]}
\title{A doubly stochastic matrices-based approach to optimal qubit routing}
\begin{document}

\maketitle
\begin{abstract}
	Swap mapping is a quantum compiler optimization that, by introducing $\mathrm{SWAP}$ gates, maps
	a logical quantum circuit to an equivalent physically implementable one.
	The physical implementability of a circuit is determined by the fulfillment of the hardware connectivity constraints.
	Therefore, the placement of the $\mathrm{SWAP}$ gates can be interpreted as a discrete optimization process.
	In this work, we employ a structure called doubly stochastic matrix, which is defined as a convex combination
	of permutation matrices. The intuition is that of making the decision process smooth.
	Doubly stochastic matrices are contained in the Birkhoff polytope, in which the vertices represent single permutation matrices.
	In essence, the algorithm uses smooth constrained optimization to slide along the edges of the polytope toward the potential
	solutions on the vertices.
	In the experiments, we show that the proposed algorithm, at the cost of additional computation time,
	can deliver significant depth reduction when compared to the state of the art algorithm SABRE.
\end{abstract}

\section{Introduction}

In the quantum computing field, the qubit routing procedure is a fundamental component of the quantum compiler. Its role is that of mapping logical qubits to
physical ones while preserving the resulting unitary (up to a permutation) and fulfilling the connectivity constraints of the target hardware.
The constraints consist of the set of pairs of physical qubits upon which a two-qubit gate can be applied.
One of the methods for solving this problem is called swap mapping \cite{maslov-qubit-placement,nannicini2021optimal}.

The routing is obtained by introducing $\mathrm{SWAP}$ gates, which determine a permutation of the mapping between logical and physical qubits.
The permutations are calculated so that the multi-qubit gates of the compiled circuit do not violate the connectivity restrictions of the hardware.
In many architectures, the $\mathrm{SWAP}$ gate is constructed using three $\mathrm{CNOT}$s, the latter,
have usually a higher error rate \cite{Magesan_2012} when compared to single-qubit gates.
Moreover, adding gates to the circuit may enlarge the circuit depth, thus increasing decoherence-related issues.
Consequently, the qubit routing is an optimization problem that not only is required to fulfill the hardware constraints, but also
to minimize some function that depends on the number of added $\mathrm{SWAP}$s and possibly the depth of the resulting circuit.
In literature this problem is found under several designations, among others, we mention the qubit allocation,
the layout synthesis and the qubit placement.

The computational complexity of the combinatorial problem behind the swap mapping has been proved in several forms.
In \cite{maslov-qubit-placement} the mapping problem is formulated as a Hamiltonian cycle problem, resulting in being NP-complete.
A reduction from the sub-graph isomorphism problem in \cite{qubit-alloc} shows that the problem is NP-hard.
This means that the design of practical algorithms should employ a heuristic component.
Among the heuristics, remarkable is the SWAP-based BidiREctional (SABRE) algorithm \cite{sabre}.
At the time of writing, the latter is considered the state of the art and is currently the default swap mapping
method for the Qiskit framework \cite{Qiskit}.


Despite the long-standing research on the topic, in \cite{Tan_2021} it was surveyed that, there is still a substantial optimality gap
on the solutions produced by the most widespread tools for circuit synthesis.

In this work we present a swap mapping algorithm based on mathematical optimization and doubly stochastic matrices (DSM) \cite{brualdi_2006}.
Doubly stochastic matrices are convex combinations of permutations matrices.
The intuition is that, since the qubit allocation makes use of swaps (permutations) we can model the decision process using a
superposition of swap and identity matrices.
Such superposition is then tuned by means of continuous parameters controlled by an optimizer. In addition, powerful algebraic properties
of the resulting cost function allow the modeling of swap count and depth minimization.
On the optimizer side, we propose a solver that scales linearly with the depth of the circuit, and our experiments show that for compiling quantum-volume circuits on 8 qubits our solver outperforms the state-of-the-art algorithm SABRE in terms of depth of the resulting circuit by 20 percentage points.


The paper is is organized as follows. In Section \ref{section:prelim} we introduce the basic definitions of the algebraic structures used throughout the formulation, Section \ref{section:formulation} contains the construction of the fundamental part of the optimization problem
--- the hardware cost function.
Sections \ref{section:optim-prob}, \ref{section:line-topo} and \ref{section:solver} expand on the optimization problem,
its key features and the numerical method.
In Section \ref{section:experiments} we report the results from our experiments: numerical evaluation of the effect of algorithm's hyper-parameters on compilation outcome, and comparisons against the state of the art algorithms. Finally the conclusions are elaborated in Section \ref{section:conclusions}.
All the proofs related to the main results can be found in the appendix.

\section{Mathematical preliminaries}
\label{section:prelim}

\begin{notation}
We introduce the basic notational conventions adopted throughout this work.
We denote with $\idenm{m}$ the $m\times m$ identity matrix, also $\onesv{m}$ denotes the vector of ones
in the real vector space $\mathbb{R}^m$.
Canonical basis vectors are identified with $\mathbf{e}_i$, also sometimes we highlight the dimension of the vector space
by denoting with $\bvec{i}{m}$, the $i$-th (starting from zero) basis vector for $\mathbb{R}^m$.
Given the relation with quantum computing we make use of the Dirac notation.
In the context of this work, we interpret $\ket{i}_m$ as the $i$-th canonical basis vector for $\mathbb{R}^m$,
that is $\ket{i}_m=\bvec{i}{m}$.
Consequently, the outer product $\ket{i}_m\bra{j}_m$ represents the rank-one matrix $\bvec{i}{m}\left(\bvec{j}{m}\right)^\top$.
For a vector $\mathbf{v} \in \mathbb{R}^n$ and some integers $i, j$ such that, $0 \le i\le j < n$
we denote with $\mathbf{v}[i\isep j]$ the sub-vector in $\mathbb{R}^{j-i+1}$ obtained through the linear transformation
\begin{align}
	\mathbf{v}[i\isep j] =& \sum_{k=0}^{m-1} \ket{k}_{m} \bra{k+i}_{n} \mathbf{v},
\end{align}
with $m=j-i+1$.
Given integers $i, j$ such that $i \le j$, we denote the set $[i, j] \cap \mathbb{Z}$ with $[i\twodots j]$.
The \textit{Hadamard product} of $n\times m$ matrices $A, B$ is denoted with $A \odot B$ and
the resulting $n\times m$ matrix has entries $(A\odot B)_{i, j}=A_{i, j} B_{i, j}$, where $A_{i, j}$
denotes the matrix entry at the $i$-th row and $j$-th column.
\end{notation}

\begin{definition}
	\label{def:dsm}
	We denote an $n \times n$ matrix $A$ with non-negative entries, as \underline{doubly stochastic} (DSM), when
	\begin{subequations}
	\begin{align}
		AJ_n =& J_nA = J_n,
	\end{align}
	where $J_n=\onesv{n}\onesv{n}^\top$ is the $n \times n$ matrix of ones.
	In other words both rows and columns of $A$ sum to 1 and $A_{i, j} \ge 0$.
	Moreover, the Birkhoff–von Neumann theorem \cite{brualdi_2006} asserts that any doubly stochastic matrix can be
	decomposed as convex combination of permutation matrices, that is
	\begin{align}
		A = \sum_{i=0}^{n! - 1} \lambda_i P_i
	\end{align}
	\end{subequations}
	with $\lambda_i \ge 0$, $\sum_i \lambda_i=1$ and $\{P_i\}$ the set of $n\times n$ permutation matrices.
\end{definition}

We mention a well known result regarding the set of DSMs.
\begin{lemma}
	\label{lemma:dsm-closed-mm}
	The set of $n\times n$ doubly stochastic matrices form a monoid\footnote{
		A monoid is defined as a set endowed with a binary associative operation
		and an identity element. In other words a monoid is a group without the inverse axiom.
		A counter-example to invertibility of DSMs is the following. Take $X$ to be any permutation matrix that swaps
		two elements, then its eigenvalues are $\{-1, 1\}$, now $\frac{1}{2}(\idenm{n}+X)$ is a DSM, but it has a zero eigenvalue,
		consequently it is singular.
	}
	under matrix multiplication.
\end{lemma}

\begin{definition}
	Given any $n \times n$ matrix $A$, we define the row-major
	\underline{vectorization} operator $\vecop: \mathcal{M}_{n, n}(K) \to K^n \otimes K^n$
	with rule
	\begin{subequations}
	\begin{align}
		\label{eq:vecr-def}
		\vecop(A)=&\sum_{j=0}^{n-1} \left(A\ket{j}_n\right) \otimes \ket{j}_n
		=\sum_{i, j=0}^{n-1} A_{i, j}\left(\ket{i}_n \otimes \ket{j}_n\right),
	\end{align}
	where $\mathcal{M}_{n, n}(K)$ is the set of $n\times n$ matrices over some field $K$.
	A special case is given by the identity matrix
	\begin{align}
		\vecop(\idenm{n}) = \sum_{i} \ket{i}_n\otimes \ket{i}_n,
	\end{align}
	\end{subequations}
	which can be interpreted as a Greenberger–Horne–Zeilinger state (GHZ) up to a scalar factor.
\end{definition}

\begin{lemma}
	\label{lemma:vec-identities}
	Let $A$ be any $n \times n$ matrix, then
	\begin{align}
		\label{eq:vec-identities}
		\vecop(A) =& (A\otimes \idenm{n}) \vecop(\idenm{n})
		= \left(\idenm{n}\otimes A^\top\right) \vecop(\idenm{n})\,.
	\end{align}
\end{lemma}
\prooflater{}
The next lemma presents some convenient identities that create a link between vectorization, tensor and Hadamard products and trace.
\begin{subequations}
\begin{lemma}
	\label{lemma:vec-tp-hp}
	Let $A, B$ be any $n \times n$ matrices, then
	\begin{align}
		\label{eq:bellt-atpb-bell}
		\vecop(\idenm{n})^\top (A\otimes B) \vecop(\idenm{n})
		=& \traceop\left(AB^\top\right)\\
		\label{eq:onest-ahpb-ones}
		=& \onesv{n}^\top (A\odot B)\onesv{n}.
	\end{align}
\end{lemma}
\end{subequations}
\prooflater{}

\section{Main results}
\subsection{The hardware cost function}
\label{section:formulation}
In this section we expand on the hardware cost function which, given a configuration of $\mathrm{SWAP}$ gates, vanishes when the composition of
$\mathrm{SWAP}$s and circuit layers fulfill the connectivity constraints.

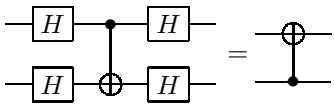
\begin{figure}
    \centering
    $\vcenter{
    \Qcircuit @C=1.0em @R=1.0em @!R {
        & \gate{H} & \ctrl{1} & \gate{H} & \qw\\
        & \gate{H} & \targ & \gate{H} & \qw
    }}\,=\vcenter{
    \Qcircuit @C=1.0em @R=1.0em @!R {
        & \targ & \qw\\
        & \ctrl{-1} & \qw
    }}$
    \caption{Reversal of the CNOT's control-target qubits.}
	\label{fig:cnotrev}
\end{figure}

Let $\mathcal{M}$ represent the hardware graph, in which the vertices are interpreted as the physical qubits and the edges the connectivity between them.
For simplicity we assume the connectivity to be undirected, also the graph $\mathcal{M}$ is assumed connected\footnote{We should also
assume that $\mathcal{M}$ is not fully-connected otherwise the swap mapping would not be necessary.} and nonempty.
In the case of quantum technologies where the only 2-qubit gate is the CNOT gate, the reversal of the control-target of such gate (Figure \ref{fig:cnotrev})
comes at no cost in terms of additional 2-qubit gates. Thus, the latter justifies the choice for the undirected graphs.
We associate to graph $\mathcal{M}$ the $m\times m$ adjacency matrix $M$.
As consequence of the assumed structure of $\mathcal{M}$, we have that $M$ is symmetric, that is $M=M^\top$.
Let $J_m=\onesv{m} \onesv{m}^\top$ be the $m\times m$ matrix of ones, so denote $M_c=J_m - M$ the adjacency matrix of the \textit{complement graph} to $\mathcal{M}$.

\begin{figure}
    \centering
	\scalebox{0.80}{$\begin{tikzpicture}
	\begin{pgfonlayer}{nodelayer}
		\node [style=none] (0) at (-10, 0) {};
		\node [style=none] (1) at (-10, 2) {};
		\node [style=none] (2) at (-10, 4) {};
		\node [style=none] (3) at (-10, -2) {};
		\node [style=none] (4) at (-10, -4) {};
		\node [style=none] (5) at (9, 4) {};
		\node [style=none] (6) at (9, 2) {};
		\node [style=none] (7) at (9, 0) {};
		\node [style=none] (8) at (9, -2) {};
		\node [style=none] (9) at (9, -4) {};
		\node [style=block, minimum width=1cm, minimum height=2.5cm] (10) at (-8, 2) {$U_{0}$};
		\node [style=block, minimum width=1cm, minimum height=1.5cm] (58) at (-8, -3) {$U_1$};
		\node [style=block, minimum width=1cm, minimum height=1.5cm] (59) at (-5, 3) {$U_2$};
		\node [style=block, minimum width=1cm, minimum height=1.5cm] (60) at (-5, -1) {$U_3$};
		\node [style=block, minimum width=1cm, minimum height=2.5cm] (61) at (-2, 0) {$U_4$};
		\node [style=block, minimum width=1cm, minimum height=2.5cm] (62) at (1, -2) {$U_5$};
		\node [style=block, minimum width=1cm, minimum height=1.5cm] (63) at (4, -1) {$U_6$};
		\node [style=block, minimum width=1cm, minimum height=4.5cm] (64) at (7, 0) {$U_7$};
		\node [style=dot1] (11) at (-9, 4) {};
		\node [style=dot1] (12) at (-7, 4) {};
		\node [style=dot1] (13) at (-9, 0) {};
		\node [style=dot1] (14) at (-7, 0) {};
		\node [style=dot1] (15) at (-9, -2) {};
		\node [style=dot1] (16) at (-7, -2) {};
		\node [style=dot1] (17) at (-9, -4) {};
		\node [style=dot1] (18) at (-7, -4) {};
		\node [style=dot1] (19) at (-6, 4) {};
		\node [style=dot1] (20) at (-4, 4) {};
		\node [style=dot1] (21) at (-6, 2) {};
		\node [style=dot1] (22) at (-4, 2) {};
		\node [style=dot1] (23) at (-6, 0) {};
		\node [style=dot1] (24) at (-4, 0) {};
		\node [style=dot1] (25) at (-6, -2) {};
		\node [style=dot1] (26) at (-4, -2) {};
		\node [style=dot1] (27) at (-3, 2) {};
		\node [style=dot1] (28) at (-1, 2) {};
		\node [style=dot1] (29) at (-3, -2) {};
		\node [style=dot1] (30) at (-1, -2) {};
		\node [style=dot1] (31) at (0, 0) {};
		\node [style=dot1] (32) at (2, 0) {};
		\node [style=dot1] (33) at (0, -4) {};
		\node [style=dot1] (34) at (2, -4) {};
		\node [style=none] (35) at (-6.5, 5) {};
		\node [style=none] (36) at (-6.5, -5) {};
		\node [style=none] (37) at (-3.5, 5) {};
		\node [style=none] (38) at (-3.5, -5) {};
		\node [style=none] (39) at (2.5, 5) {};
		\node [style=none] (40) at (2.5, -5) {};
		\node [style=dot1] (41) at (3, 0) {};
		\node [style=dot1] (42) at (5, 0) {};
		\node [style=dot1] (43) at (3, -2) {};
		\node [style=dot1] (44) at (5, -2) {};
		\node [style=dot1] (45) at (6, 4) {};
		\node [style=dot1] (46) at (8, 4) {};
		\node [style=dot1] (47) at (6, -4) {};
		\node [style=dot1] (48) at (8, -4) {};
		\node [style=none] (49) at (8.5, 5) {};
		\node [style=none] (50) at (8.5, -5) {};
		\node [style=none] (51) at (-9, 2) {};
		\node [style=none] (52) at (-7, 2) {};
		\node [style=none] (53) at (-10.5, 4) {$q_4$};
		\node [style=none] (54) at (-10.5, 2) {$q_3$};
		\node [style=none] (55) at (-10.5, 0) {$q_2$};
		\node [style=none] (56) at (-10.5, -2) {$q_1$};
		\node [style=none] (57) at (-10.5, -4) {$q_0$};
		\node [style=none] (65) at (6, 2) {};
		\node [style=none] (66) at (6, 0) {};
		\node [style=none] (67) at (6, -2) {};
		\node [style=none] (68) at (8, 2) {};
		\node [style=none] (69) at (8, 0) {};
		\node [style=none] (70) at (8, -2) {};
		\node [style=none] (71) at (0, -2) {};
		\node [style=none] (72) at (2, -2) {};
		\node [style=none] (73) at (-1, 0) {};
		\node [style=none] (74) at (-3, 0) {};
	\end{pgfonlayer}
	\begin{pgfonlayer}{edgelayer}
		\draw [style=ldashed] (35.center) to (36.center);
		\draw [style=ldashed] (37.center) to (38.center);
		\draw [style=ldashed] (39.center) to (40.center);
		\draw [style=ldashed] (49.center) to (50.center);
		\draw [style=wire] (1.center) to (51.center);
		\draw [style=wire] (52.center) to (21);
		\draw [style=wire] (22) to (27);
		\draw [style=wire] (2.center) to (11);
		\draw [style=wire] (12) to (19);
		\draw [style=wire] (20) to (45);
		\draw [style=wire] (46) to (5.center);
		\draw [style=wire] (68.center) to (6.center);
		\draw [style=wire] (28) to (65.center);
		\draw [style=wire] (0.center) to (13);
		\draw [style=wire] (14) to (23);
		\draw [style=wire] (24) to (74.center);
		\draw [style=wire] (73.center) to (31);
		\draw [style=wire] (32) to (41);
		\draw [style=wire] (42) to (66.center);
		\draw [style=wire] (69.center) to (7.center);
		\draw [style=wire] (3.center) to (15);
		\draw [style=wire] (16) to (25);
		\draw [style=wire] (26) to (29);
		\draw [style=wire] (30) to (71.center);
		\draw [style=wire] (72.center) to (43);
		\draw [style=wire] (44) to (67.center);
		\draw [style=wire] (70.center) to (8.center);
		\draw [style=wire] (4.center) to (17);
		\draw [style=wire] (18) to (33);
		\draw [style=wire] (34) to (47);
		\draw [style=wire] (48) to (9.center);
	\end{pgfonlayer}
\end{tikzpicture}$}
    \caption{Example of commutative layering for a circuit of 2-qubit blocks.
	{The blocks represent generic $U(4)$ unitaries where the dots identify the qubits upon with
	each unitary acts. The vertical dashed lines determine the layering, so for example
	take $U_0$ and $U_1$ from the first layer, then
	$(U_0 \otimes I^{\otimes 2}) (I^{\otimes 2} \otimes U_1) = (I^{\otimes 2} \otimes U_1) (U_0 \otimes I^{\otimes 2})$,
	that is $U_0, U_1$ commute.}}
	\label{fig:comm-blocks}
\end{figure}
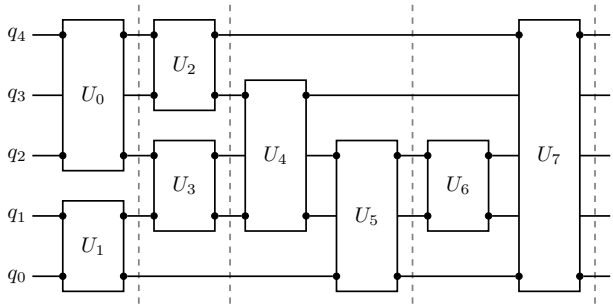

\begin{figure}
    \centering
	\scalebox{0.80}{$\begin{tikzpicture}
	\begin{pgfonlayer}{nodelayer}
		\node [style=none] (0) at (-5.25, 0) {};
		\node [style=none] (1) at (-5.25, 2) {};
		\node [style=none] (2) at (-5.25, 4) {};
		\node [style=none] (3) at (-5.25, -2) {};
		\node [style=none] (4) at (-5.25, -4) {};
		\node [style=none] (5) at (5.75, 4) {};
		\node [style=none] (6) at (5.75, 2) {};
		\node [style=none] (7) at (5.75, 0) {};
		\node [style=none] (8) at (5.75, -2) {};
		\node [style=none] (9) at (5.75, -4) {};
		\node [style=dot] (15) at (-4.25, -2) {};
		\node [style=dot] (17) at (-4.25, -4) {};
		\node [style=dot] (19) at (-2.25, 4) {};
		\node [style=dot] (23) at (-2.25, 0) {};
		\node [style=dot] (25) at (-2.25, -2) {};
		\node [style=dot] (27) at (-0.25, 2) {};
		\node [style=dot] (29) at (-0.25, -2) {};
		\node [style=dot] (31) at (0.75, 0) {};
		\node [style=dot] (33) at (0.75, -4) {};
		\node [style=none] (35) at (-3.25, 5) {};
		\node [style=none] (36) at (-3.25, -5) {};
		\node [style=none] (37) at (-1.25, 5) {};
		\node [style=none] (38) at (-1.25, -5) {};
		\node [style=none] (39) at (1.75, 5) {};
		\node [style=none] (40) at (1.75, -5) {};
		\node [style=dot] (41) at (2.75, 0) {};
		\node [style=dot] (43) at (2.75, -2) {};
		\node [style=dot] (45) at (3.75, 4) {};
		\node [style=dot] (47) at (3.75, -4) {};
		\node [style=none] (49) at (4.75, 5) {};
		\node [style=none] (50) at (4.75, -5) {};
		\node [style=none] (53) at (-5.75, 4) {$q_4$};
		\node [style=none] (54) at (-5.75, 2) {$q_3$};
		\node [style=none] (55) at (-5.75, 0) {$q_2$};
		\node [style=none] (56) at (-5.75, -2) {$q_1$};
		\node [style=none] (57) at (-5.75, -4) {$q_0$};
		\node [style=none] (59) at (-2.75, 3) {$U_2$};
		\node [style=none] (60) at (-2.75, -1) {$U_3$};
		\node [style=none] (61) at (-0.75, 1) {$U_4$};
		\node [style=none] (62) at (0.25, -3) {$U_5$};
		\node [style=none] (63) at (2.25, -1) {$U_6$};
		\node [style=none] (64) at (3.25, 3) {$U_7$};
		\node [style=dot] (75) at (-4.25, 4) {};
		\node [style=dot] (76) at (-4.25, 0) {};
		\node [style=dot] (77) at (-2.25, 2) {};
		\node [style=none] (78) at (-4.75, 3) {$U_0$};
		\node [style=none] (79) at (-4.75, -3) {$U_1$};
	\end{pgfonlayer}
	\begin{pgfonlayer}{edgelayer}
		\draw [style=ldashed] (35.center) to (36.center);
		\draw [style=ldashed] (38.center) to (37.center);
		\draw [style=ldashed] (39.center) to (40.center);
		\draw [style=ldashed] (49.center) to (50.center);
		\draw [style=wire] (2.center) to (5.center);
		\draw [style=wire] (1.center) to (6.center);
		\draw [style=wire] (0.center) to (7.center);
		\draw [style=wire] (3.center) to (8.center);
		\draw [style=wire] (4.center) to (9.center);
		\draw [style=wirett] (75) to (76);
		\draw [style=wirett] (15) to (17);
		\draw [style=wirett] (19) to (77);
		\draw [style=wirett] (23) to (25);
		\draw [style=wirett] (27) to (29);
		\draw [style=wirett] (31) to (33);
		\draw [style=wirett] (41) to (43);
		\draw [style=wirett] (45) to (47);
	\end{pgfonlayer}
\end{tikzpicture} \quad\mapsto\quad \begin{tikzpicture}
	\begin{pgfonlayer}{nodelayer}
		\node [style=none] (0) at (-6, 0) {};
		\node [style=none] (1) at (-6, 2) {};
		\node [style=none] (2) at (-6, 4) {};
		\node [style=none] (3) at (-6, -2) {};
		\node [style=none] (4) at (-6, -4) {};
		\node [style=none] (5) at (4, 4) {};
		\node [style=none] (6) at (4, 2) {};
		\node [style=none] (7) at (4, 0) {};
		\node [style=none] (8) at (4, -2) {};
		\node [style=none] (9) at (4, -4) {};
		\node [style=dot] (15) at (-4, -2) {};
		\node [style=dot] (17) at (-4, -4) {};
		\node [style=dot] (19) at (-2, 4) {};
		\node [style=dot] (23) at (-2, 0) {};
		\node [style=dot] (25) at (-2, -2) {};
		\node [style=dot] (27) at (0, 2) {};
		\node [style=dot] (29) at (0, 4) {};
		\node [style=dot] (31) at (0, -2) {};
		\node [style=dot] (33) at (0, -4) {};
		\node [style=none] (35) at (-3, 5) {};
		\node [style=none] (36) at (-3, -5) {};
		\node [style=none] (37) at (-1, 5) {};
		\node [style=none] (38) at (-1, -5) {};
		\node [style=none] (39) at (1, 5) {};
		\node [style=none] (40) at (1, -5) {};
		\node [style=dot] (41) at (2, -2) {};
		\node [style=dot] (43) at (2, -4) {};
		\node [style=dot] (45) at (2, 2) {};
		\node [style=dot] (47) at (2, 0) {};
		\node [style=none] (49) at (3, 5) {};
		\node [style=none] (50) at (3, -5) {};
		\node [style=none] (53) at (-6.5, 4) {$q_4$};
		\node [style=none] (54) at (-6.5, 2) {$q_3$};
		\node [style=none] (55) at (-6.5, 0) {$q_2$};
		\node [style=none] (56) at (-6.5, -2) {$q_1$};
		\node [style=none] (57) at (-6.5, -4) {$q_0$};
		\node [style=none] (59) at (-2.5, 3) {$U_2$};
		\node [style=none] (60) at (-1.5, -1) {$U_3$};
		\node [style=none] (61) at (-0.5, 3) {$U_4$};
		\node [style=none] (62) at (-0.5, -3) {$U_5$};
		\node [style=none] (63) at (2.5, 1) {$U_6$};
		\node [style=none] (64) at (2.5, -3) {$U_7$};
		\node [style=dot] (75) at (-4, 2) {};
		\node [style=dot] (76) at (-4, 0) {};
		\node [style=dot] (77) at (-2, 2) {};
		\node [style=none] (78) at (-4.5, 1) {$U_0$};
		\node [style=none] (79) at (-4.5, -3) {$U_1$};
		\node [style=none] (80) at (-4, 4) {};
		\node [style=none] (81) at (0, 0) {};
		\node [style=none] (53) at (4.5, 4) {$q_3$};
		\node [style=none] (54) at (4.5, 2) {$q_1$};
		\node [style=none] (55) at (4.5, 0) {$q_2$};
		\node [style=none] (56) at (4.5, -2) {$q_4$};
		\node [style=none] (57) at (4.5, -4) {$q_0$};
	\end{pgfonlayer}
	\begin{pgfonlayer}{edgelayer}
		\draw [style=ldashed] (35.center) to (36.center);
		\draw [style=ldashed] (38.center) to (37.center);
		\draw [style=ldashed] (39.center) to (40.center);
		\draw [style=ldashed] (49.center) to (50.center);
		\draw [style=wirett] (75) to (76);
		\draw [style=wirett] (15) to (17);
		\draw [style=wirett] (19) to (77);
		\draw [style=wirett] (23) to (25);
		\draw [style=wirett] (27) to (29);
		\draw [style=wirett] (31) to (33);
		\draw [style=wirett] (41) to (43);
		\draw [style=wirett] (45) to (47);
		\draw [style=wire, in=180, out=0] (2.center) to (75);
		\draw [style=wire, in=180, out=0] (1.center) to (80.center);
		\draw [style=wire] (80.center) to (5.center);
		\draw [style=wire] (0.center) to (76);
		\draw [style=wire] (3.center) to (15);
		\draw [style=wire] (4.center) to (9.center);
		\draw [style=wire, in=180, out=0] (76) to (25);
		\draw [style=wire, in=-180, out=0] (15) to (23);
		\draw [style=wire] (75) to (77);
		\draw [style=wire, in=180, out=0] (77) to (81.center);
		\draw [style=wire, in=180, out=0] (23) to (27);
		\draw [style=wire] (25) to (31);
		\draw [style=wire, in=180, out=0] (81.center) to (41);
		\draw [style=wire, in=-180, out=0] (31) to (47);
		\draw [style=wire] (27) to (6.center);
		\draw [style=wire] (41) to (8.center);
		\draw [style=wire] (47) to (7.center);
	\end{pgfonlayer}
\end{tikzpicture}$}
    \caption{The braid diagram with gate arcs (left) corresponding to the circuit in Figure \ref{fig:comm-blocks}.
	{In this example we assume the line connectivity, so on the right we propose a qubit allocation solution.
	Note on the right-hand side that the initial and final qubit layouts do not match.
	Here the braids on the right-hand side can be interpreted as the permutation of logical qubits ($q_k$) associated to
	physical ones (vertical positions).}}
	\label{fig:comm-blocks-braid}
\end{figure}
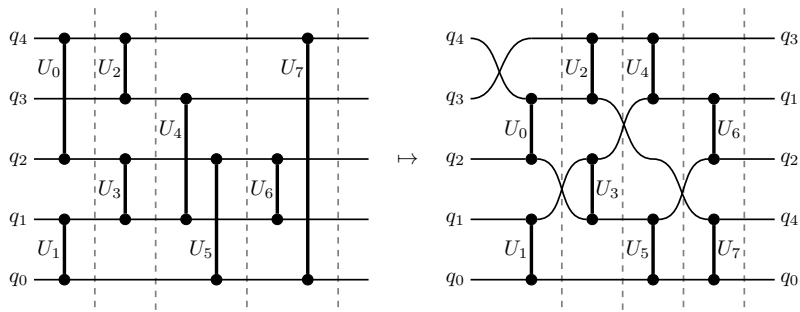

As outlined in \cite{nannicini2021optimal}, we decompose the circuit into layers of commuting two-qubits gates.
An example of such circuit partitioning is presented in Figure \ref{fig:comm-blocks}.
The same circuit is also presented in Figure \ref{fig:comm-blocks-braid} using a diagrammatic form that capture the essential information
required for the swap mapping algorithm.
Let $\timeent{\mathcal{G}}$ denote the graph corresponding to the layer at time step $t \in \intset{0}{T-1}$, and $\timeent{G}$ its adjacency matrix.
The vertices of $\timeent{\mathcal{G}}$ correspond to the logical qubits, and the edges are the two qubit gates.
Let $V_{\timeent{\mathcal{G}}}$ denote the set of vertices of the graph $\timeent{\mathcal{G}}$, then we impose $|V_\mathcal{M}|=|V_{\timeent{\mathcal{G}}}|=m$ and
$V_{\timeent[t_1]{\mathcal{G}}}=V_{\timeent[t_2]{\mathcal{G}}}$ for all $t_1, t_2 \in \intset{0}{T-1}$.

Now, let $P$ be an $m \times m$ permutation matrix, then we devise the following cost function for layer $t$,
\begin{align}
	\label{eq:cost:first}
	\timeent{\ell}(P) =& \onesv{m}^\top \left[\left(P\timeent{G}P^\top\right) \odot M_c\right] \onesv{m} \ge 0.
\end{align}
We observe that the global minima (also matching with zeros) of $\timeent{\ell}$ correspond to the permutations of the logical qubits
such that the two-qubits gates overlap hardware arcs. In general the solution $P$ to the problem of making $\timeent{\ell}(P)$ vanish
is not unique, however later we will add a further restriction, that is the minimization of the number of $\mathrm{SWAP}$s generating $P$.

We obtain an equivalent formulation for the cost function in \eqref{eq:cost:first}.
Using Lemmas \ref{lemma:vec-identities} and \ref{lemma:vec-tp-hp} we have
\begin{subequations}
\begin{align}
	\timeent{\ell}(P)
	=& \traceop\left(\left(P\timeent{G} P^\top\right) M_c^\top\right)\\
	=& \traceop\left(P\timeent{G} \left(M_c P\right)^\top\right)\\
	\label{eq:cost:tp-intermediate-1}
	\overset{\eqref{eq:bellt-atpb-bell}}{=}& \vecop(\idenm{m})^\top
	\left(P\timeent{G}\right)\otimes\left(M_c P\right)
	\vecop(\idenm{m})\\
	=& \vecop(\idenm{m})^\top
	\left(\idenm{m} \otimes M_c\right)
	\left(P \otimes P\right)
	\left(\timeent{G} \otimes \idenm{m}\right)
	\vecop(\idenm{m})\\
	\label{eq:cost:tp}
	\overset{\eqref{eq:vec-identities}}{=}& \vecop(M_c)^\top \left(P\otimes P\right) \vecop(\timeent{G}),
\end{align}
\end{subequations}
The new form in \eqref{eq:cost:tp} is convenient for the theory that follows, the key fact is that the cost function is now \underline{linear}
with respect to $P\otimes P$.
The intent is that of obtaining a cost value as a superposition of cost values (non-negative) for individual solutions. Specifically, each
solution is determined by a permutation matrix $P_i$, then given a set of candidate solutions $\{P_i\}$ we obtain the convex combination
\begin{subequations}
\begin{align}
	\label{eq:cost:first-with-lambda}
	\sum_i \lambda_i \timeent{\ell}(P_i) =& \sum_i \lambda_i \vecop(M_c)^\top \left(P_i\otimes P_i\right) \vecop(\timeent{G})\\
	\label{eq:first-entangl-dsm}
	=& \vecop(M_c)^\top \left(\sum_i \lambda_i \left(P_i\otimes P_i\right)\right) \vecop(\timeent{G}),
\end{align}
\end{subequations}
with $\boldsymbol{\lambda} \in \Delta_S$ where $\Delta_S$ is the \textit{unit simplex} in $S$ dimensions \cite{BeckBook}.
In \eqref{eq:first-entangl-dsm} we have the first appearance of an interesting structure, whose characterization is given
by the following lemma.
\begin{lemma}
	\label{lemma:dsm-p-tp-p}
	Let $Q=\sum_i \lambda_i P_i$ be a doubly stochastic matrix, where $P_i$ are $m\times m$
	permutation matrices. Then
	\begin{align}
		K=&\sum_i \lambda_i P_i \otimes P_i
	\end{align}
	is doubly stochastic.
\end{lemma}
\prooflater{}

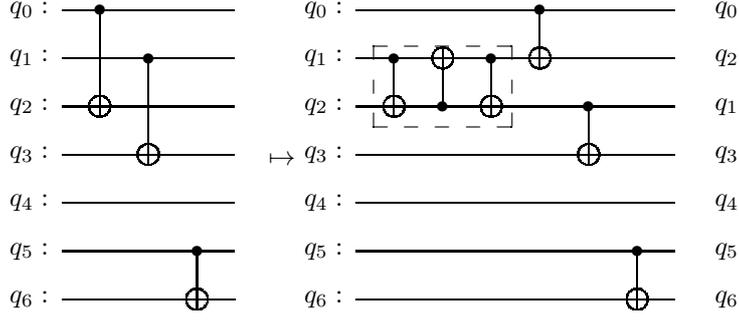
\begin{figure}
    \centering
    $\vcenter{\Qcircuit @C=1.0em @R=1.0em @!R {
        & \lstick{q_0:} & \ctrl{2} & \qw      & \qw & \qw\\
        & \lstick{q_1:} & \qw      & \ctrl{2} & \qw & \qw\\
        & \lstick{q_2:} & \targ    & \qw      & \qw & \qw\\
        & \lstick{q_3:} & \qw      & \targ    & \qw & \qw\\
        & \lstick{q_4:} & \qw      & \qw      & \qw & \qw\\
        & \lstick{q_5:} & \qw      & \qw      & \ctrl{1} & \qw\\
        & \lstick{q_6:} & \qw      & \qw      & \targ & \qw
    }}\quad\mapsto\quad
	\vcenter{\Qcircuit @C=1.0em @R=1.0em @!R {
        & \lstick{q_0:} & \qw      & \qw       & \qw      & \ctrl{1} & \qw      & \qw & \qw & \rstick{q_0}\\
        & \lstick{q_1:} & \ctrl{1} & \targ     & \ctrl{1} & \targ    & \qw      & \qw & \qw & \rstick{q_2}\\
        & \lstick{q_2:} & \targ    & \ctrl{-1} & \targ    & \qw      & \ctrl{1} & \qw & \qw & \rstick{q_1}\\
        & \lstick{q_3:} & \qw      & \qw       & \qw      & \qw      & \targ    & \qw & \qw & \rstick{q_3}\\
        & \lstick{q_4:} & \qw      & \qw       & \qw      & \qw      & \qw      & \qw & \qw & \rstick{q_4}\\
        & \lstick{q_5:} & \qw      & \qw       & \qw      & \qw      & \qw      & \ctrl{1} & \qw & \rstick{q_5}\\
        & \lstick{q_6:} & \qw      & \qw       & \qw      & \qw      & \qw      & \targ & \qw & \rstick{q_6}
		\gategroup{2}{3}{3}{5}{.7em}{--}
    }}$
    \caption{On the left-hand side, a circuit layer corresponding to the adjacency matrix $\timeent{G}$
	considered in Figure \ref{fig:simplei}. Note that the CNOTs commute.
	On the right-hand side, the circuit after the application of the swap (highlighted).}
	\label{fig:circ:layer}
\end{figure}

The function $\timeent{\ell}(P)$ is the key component of what will be defined later as the \textit{hardware cost function}.
In Figure \ref{fig:simplei} we provide a visual example of the cost function machinery, also Figure \ref{fig:circ:layer}
presents the circuit related to the example matrices.
\begin{remark}
	We highlight again the linearity of the optimization problem determined by \eqref{eq:cost:first-with-lambda}.
	Take for simplicity the problem of minimizing the hardware cost function for the layer $t$.
	Consider $m$ qubits and fix the set of permutation matrices $\{P_i\}$ corresponding to the representations of the
	elements of the symmetric group $\symmg{m}$.
	Then the problem takes the form
	\begin{subequations}
	\begin{align}
		\underset{\boldsymbol{\lambda} \in \mathbb{R}^{m!}}{\mathrm{min}}\,&\sum_{i=0}^{m! - 1} \lambda_i \timeent{\ell}(P_i),\\
		\text{s.t.}\,& \lambda_i \ge 0\qquad \forall\, i,\\
		& \sum_i \lambda_i = 1.
	\end{align}
	\end{subequations}
	The latter is clearly a linear problem over a convex set,
	however the difficulty arises from the cardinality $m!$ of the set of permutations.
	We avoid this difficulty by introducing a construction that produces a subgroup of
	permutations controlled by a polynomial number of parameters (w.r.t. $m$).

	\myendofremark{}
\end{remark}

\begin{figure}
	\centering
	\begin{align*}
		M=
		\begin{pmatrix}
			\bullet&\bullet&\cdot&\cdot&\cdot&\cdot&\cdot\\
			\bullet&\bullet&\bullet&\cdot&\cdot&\cdot&\cdot\\
			\cdot&\bullet&\bullet&\bullet&\cdot&\cdot&\cdot\\
			\cdot&\cdot&\bullet&\bullet&\bullet&\cdot&\cdot\\
			\cdot&\cdot&\cdot&\bullet&\bullet&\bullet&\cdot\\
			\cdot&\cdot&\cdot&\cdot&\bullet&\bullet&\bullet\\
			\cdot&\cdot&\cdot&\cdot&\cdot&\bullet&\bullet
		\end{pmatrix},
		&
		\quad
		\timeent{G}=
		\begin{pmatrix}
			\cdot&\cdot&\bullet&\cdot&\cdot&\cdot&\cdot\\
			\cdot&\cdot&\cdot&\bullet&\cdot&\cdot&\cdot\\
			\bullet&\cdot&\cdot&\cdot&\cdot&\cdot&\cdot\\
			\cdot&\bullet&\cdot&\cdot&\cdot&\cdot&\cdot\\
			\cdot&\cdot&\cdot&\cdot&\cdot&\cdot&\cdot\\
			\cdot&\cdot&\cdot&\cdot&\cdot&\cdot&\bullet\\
			\cdot&\cdot&\cdot&\cdot&\cdot&\bullet&\cdot
		\end{pmatrix},\\
		P=\begin{pmatrix}
			\bullet&\cdot&\cdot&\cdot&\cdot&\cdot&\cdot\\
			\cdot&\cdot&\bullet&\cdot&\cdot&\cdot&\cdot\\
			\cdot&\bullet&\cdot&\cdot&\cdot&\cdot&\cdot\\
			\cdot&\cdot&\cdot&\bullet&\cdot&\cdot&\cdot\\
			\cdot&\cdot&\cdot&\cdot&\bullet&\cdot&\cdot\\
			\cdot&\cdot&\cdot&\cdot&\cdot&\bullet&\cdot\\
			\cdot&\cdot&\cdot&\cdot&\cdot&\cdot&\bullet
		\end{pmatrix},
		&
		\quad
		P\timeent{G}P^\top=
		\begin{pmatrix}
			\cdot&\bullet&\cdot&\cdot&\cdot&\cdot&\cdot\\
			\bullet&\cdot&\cdot&\cdot&\cdot&\cdot&\cdot\\
			\cdot&\cdot&\cdot&\bullet&\cdot&\cdot&\cdot\\
			\cdot&\cdot&\bullet&\cdot&\cdot&\cdot&\cdot\\
			\cdot&\cdot&\cdot&\cdot&\cdot&\cdot&\cdot\\
			\cdot&\cdot&\cdot&\cdot&\cdot&\cdot&\bullet\\
			\cdot&\cdot&\cdot&\cdot&\cdot&\bullet&\cdot
		\end{pmatrix}\,.
	\end{align*}
	\caption{In this example we illustrate the matrices involved in \eqref{eq:cost:first}.
	We define graphically with $\cdot, \bullet$ the values $0, 1$ respectively.
	In the circuit adjacency matrix $\timeent{G}$ (corresponding to left part of Figure \ref{fig:circ:layer}) we see that some of the
	non-zero entries do not overlap with the hardware topology $M$. However, after applying the permutation $P$
	we see that the total overlap is established in $P\timeent{G}P^\top$, hence $\timeent{\ell}(P)$ vanishes.}
	\label{fig:simplei}
\end{figure}

The machinery for the swapping requires consistency between layers, that is, we have to assure the connectivity of logical qubits as we pass from one layer to the next.
Let $\timeent{P}$ be the permutation applied before the layer at time step $t$, for all $t \in \timestepsset{}$.
We define the finite sequence $\left(\timeent{C}\right)_{t \in \timestepsset{}}$ of permutations composed up to time $t$ as
\begin{subequations}
\begin{align}
	\timeent[0]{C} =& \timeent[0]{P},\\
	\label{eq:perm-comp-time-t}
	\timeent{C} =& \timeent{P} \timeent[t-1]{C},\quad 1 \le t \le T - 1,
\end{align}
\end{subequations}
expanded, the sequence takes the following form
\begin{align*}
	\timeent[0]{C} =& \timeent[0]{P},\\
	\timeent[1]{C} =& \timeent[1]{P} \timeent[0]{P},\\
	&\vdots\\
	\timeent[T - 1]{C} =& \timeent[T - 1]{P} \cdots \timeent[1]{P} \timeent[0]{P}\,.
\end{align*}

Given a sequence of permutations $\left(\timeent{P}\right)_{t \in \timestepsset{}}$ we obtain the \textit{hardware cost function} for the overall circuit, so
\begin{align}
	\label{eq:circcostfun}
	\mathcal{L}\left(\,(\timeent{P})_t\,\right) =& \sum_{t=0}^{T-1} \timeent{\ell}\left(\timeent{C}\right),
	\mynotenonum{Note that the sequence $\left(\timeent{C}\right)_t$ depends on $\left(\timeent{P}\right)_t$.}
\end{align}
again it can be shown that the global minima (zeros) correspond to the set of permutations that implements the circuit on allowable hardware arcs
(the edges of graph $\mathcal{M}$).
In relation to Figure \ref{fig:comm-blocks-braid}, the permutations $\timeent{P}$ correspond to the braids, and the graphs $\timeent{\mathcal{G}}$
match the layers with the gate arcs implementing the edges.

The limited connectivity of the hardware structure\footnote{The hardware graph $\mathcal{M}$ is assumed to be not fully-connected.}
implies that the permutations $\timeent{P}$ must be generated by the swaps corresponding to the directly connected vertices of graph $\mathcal{M}$.
However, before expanding the aforementioned restriction we first proceed with the definition of $\mathrm{SWAP}$.
Let $m$ be the number of qubits of the circuit and assume $m \ge 2$.
We define the swap operator w.r.t. distinct vertices $i, j \in \intset{0}{m-1}$, as \newcommand{\swapop}[1]{\mathrm{SWAP}_{#1}}
\begin{align}
	\label{eq:swap-def}
	\swapop{m}(i, j) \coloneqq& \idenm{m} - \ket{i}_m\bra{i}_m - \ket{j}_m\bra{j}_m
	+ \ket{i}_m\bra{j}_m + \ket{j}_m\bra{i}_m,
\end{align}
equivalently the operator is determined by its action on the vectors $\ket{k}_m$ with $k\in\intset{0}{m-1}$,
\begin{align*}
	\swapop{m}(i, j) \ket{k}_m =& \begin{cases}
		\ket{j}_m, & k=i,\\
		\ket{i}_m, & k=j,\\
		\ket{k}_m, & \text{otherwise}\,.
	\end{cases}
\end{align*}

\begin{remark}
	The $\swapop{m}$ defined in \eqref{eq:swap-def} should be considered as the classical counterpart of the swap operator $\swapop{m}^Q$
	in the context of quantum circuits.
	The $\swapop{m}(i, j)$ operator acts on a space having dimension $m$,
	also the result of its action is the swap of the the basis vectors $\ket{i}_m$ and $\ket{j}_m$.
	The $\swapop{m}^Q(i, j)$ instead, acts on a space of dimension $2^m$ and its action swaps $2^{m-1}$ basis vectors.
	The latter is represented diagrammatically using the symbol
	$\vcenter{\Qcircuit @C=1.0em @R=1.0em{& \qswap & \qw\\ & \qswap \qwx & \qw}}\,$,
	however in the context of this work,
	the same symbol is used to indicate both operators interchangeably.

	\myendofremark{}
\end{remark}

We obtain the set of generators for the permutations implementing the swap mapping, so
\begin{align}
	\label{eq:generating-swaps-set}
	\mathcal{P}_{\mathcal{M}} =& \left\{\swapop{m}(i, j) \middle| M_{i, j}=1, i<j\right\},
\end{align}
where $M$ is the adjacency matrix for the hardware couplings.
In other words, each permutation $\timeent{P}$ for layer $t$, is constructed by composing a subset of elements of
$\mathcal{P}_{\mathcal{M}}$.
Later we will introduce an objective that aims at minimizing the number of elements of such construction.

We extend again the approach by introducing a sort of 'smooth swap', that is a swap operator in superposition with the identity.
We define the \textit{smooth swap} operator $\mathrm{SSWAP}_m(i, j, \theta)$ as one of the following equivalent forms
\begin{subequations}
\begin{align}
	\label{eq:sswap-def-cos-sin}
	\mathrm{SSWAP}_m(i, j, \theta) \coloneqq& \cos^2(\theta) \idenm{m} + \sin^2(\theta) \swapop{m}(i, j)\\
	=& \idenm{m} + \sin^2(\theta) \left(\swapop{m}(i, j) - \idenm{m}\right)\\
	=& \idenm{m} + \sin^2(\theta) \left(
	\ket{i}_m\bra{j}_m + \ket{j}_m\bra{i}_m - \ket{i}_m\bra{i}_m - \ket{j}_m\bra{j}_m\right),
\end{align}
\end{subequations}
which can be shown to be \textit{convex combinations} of the identity matrix and the swap acting on vertices $i, j$.
It follows from the Birkhoff–von Neumann theorem that $\mathrm{SSWAP}_m(i, j, \theta)$ is a \textit{doubly stochastic matrix} \cite{matanalysis}.
The specific structure of the matrix $\mathrm{SSWAP}_m(i, j, \theta)$ is also known as an \textit{elementary doubly stochastic matrix} \cite{brualdi_2006}.
By Lemma \ref{lemma:dsm-closed-mm}, the set of DSM of the same size is closed under matrix multiplication, so by composing multiple $\mathrm{SSWAP}$ operators we
obtain another matrix of the same class.
Notably, in our case, when the $\theta$s are integer multiples of $\frac{\pi}{2}$ we obtain a vertex of the \textit{Birkhoff polytope}, that is a permutation
matrix. This means that we can substitute the component $P \otimes P$ in \eqref{eq:cost:tp} with a composition (depending on some hyper-parameters) of operators of
the form,
\begin{subequations}
\begin{align}
	\label{eq:psswap-def-cos-sin}
	\mathrm{PSSWAP}_m(i, j, \theta)
	\coloneqq& \cos^2(\theta) \idenm{m}^{\otimes 2} + \sin^2(\theta) \swapop{m}(i, j)^{\otimes 2}\\
	\label{eq:psswapdef}
	=& \idenm{m}^{\otimes 2} + \sin^2(\theta) \left(\swapop{m}(i, j)^{\otimes 2} - \idenm{m}^{\otimes 2}\right)\\
	=& \frac{1}{2}\left(\idenm{m}^{\otimes 2} + \swapop{m}(i, j)^{\otimes 2}\right)
	+ \frac{\cos(2\theta)}{2} \left(\idenm{m}^{\otimes 2} - \swapop{m}(i, j)^{\otimes 2}\right)\,.
\end{align}
\end{subequations}
Note that here we extended the concept of $\mathrm{SSWAP}$ to that of $\mathrm{PSSWAP}$, where the prefix $\mathrm{P}$ stands for 'parallelized'
which resembles the effect of the tensor power.
In the next section we continue with the substitution of the $\mathrm{PSSWAP}$ in the cost function \eqref{eq:circcostfun},
resulting in the weighted sum of costs corresponding to the superposition of solutions determined by the parameters $\theta$s.

\begin{remark}
	In relation to the definition of $\mathrm{PSSWAP}$, we highlight that in general
	\begin{align}
		\mathrm{PSSWAP}(i, j, \theta)\ne \mathrm{SSWAP}_m(i, j, \theta)^{\otimes 2}\,.
	\end{align}

	\myendofremark{}
\end{remark}



\begin{figure}
    \centering
	\scalebox{0.80}{$\begin{tikzpicture}
	\begin{pgfonlayer}{nodelayer}
		\node [style=none] (0) at (-9, 0) {};
		\node [style=none] (1) at (-9, 2) {};
		\node [style=none] (2) at (-9, 4) {};
		\node [style=none] (3) at (-9, -2) {};
		\node [style=none] (4) at (-9, -4) {};
		\node [style=none] (5) at (8, 4) {};
		\node [style=none] (6) at (8, 2) {};
		\node [style=none] (7) at (8, 0) {};
		\node [style=none] (9) at (8, -4) {};
		\node [style=none] (53) at (-9.5, 4) {$q_4$};
		\node [style=none] (54) at (-9.5, 2) {$q_3$};
		\node [style=none] (55) at (-9.5, 0) {$q_2$};
		\node [style=none] (56) at (-9.5, -2) {$q_1$};
		\node [style=none] (57) at (-9.5, -4) {$q_0$};
		\node [style=block, minimum width=1cm, minimum height=1.5cm] (10) at (-6, 1) {$U_{0}$};
		\node [style=block, minimum width=1cm, minimum height=1.5cm] (58) at (-6, -3) {$U_1$};
		\node [style=block, minimum width=1cm, minimum height=1.5cm] (59) at (-2, 3) {$U_2$};
		\node [style=block, minimum width=1cm, minimum height=1.5cm] (60) at (-2, -1) {$U_3$};
		\node [style=block, minimum width=1cm, minimum height=1.5cm] (61) at (2, 3) {$U_4$};
		\node [style=block, minimum width=1cm, minimum height=1.5cm] (62) at (2, -3) {$U_5$};
		\node [style=block, minimum width=1cm, minimum height=1.5cm] (63) at (6, 1) {$U_6$};
		\node [style=block, minimum width=1cm, minimum height=1.5cm] (64) at (6, -3) {$U_7$};
		\node [style=none] (70) at (8, -2) {};
		\node [style=dot1] (13) at (-7, 0) {};
		\node [style=dot1] (14) at (-5, 0) {};
		\node [style=dot1] (15) at (-7, -2) {};
		\node [style=dot1] (16) at (-5, -2) {};
		\node [style=dot1] (17) at (-7, -4) {};
		\node [style=dot1] (18) at (-5, -4) {};
		\node [style=dot1] (19) at (-3, 4) {};
		\node [style=dot1] (20) at (-1, 4) {};
		\node [style=dot1] (21) at (-3, 2) {};
		\node [style=dot1] (22) at (-1, 2) {};
		\node [style=dot1] (23) at (-3, 0) {};
		\node [style=dot1] (24) at (-1, 0) {};
		\node [style=dot1] (26) at (1, -2) {};
		\node [style=dot1] (27) at (1, 2) {};
		\node [style=dot1] (75) at (-7, 2) {};
		\node [style=dot1] (76) at (-5, 2) {};
		\node [style=dot1] (77) at (1, -4) {};
		\node [style=dot1] (78) at (3, -4) {};
		\node [style=dot1] (79) at (1, 4) {};
		\node [style=none] (82) at (-8, 4) {$\times$};
		\node [style=none] (83) at (-8, 2) {$\times$};
		\node [style=none] (84) at (-4, 0) {$\times$};
		\node [style=none] (85) at (-4, -2) {$\times$};
		\node [style=dot1] (86) at (3, 4) {};
		\node [style=dot1] (87) at (3, 2) {};
		\node [style=dot1] (88) at (3, -2) {};
		\node [style=dot1] (89) at (-3, -2) {};
		\node [style=dot1] (90) at (-1, -2) {};
		\node [style=none] (91) at (0, 2) {$\times$};
		\node [style=none] (92) at (0, 0) {$\times$};
		\node [style=dot1] (93) at (5, 2) {};
		\node [style=dot1] (94) at (7, 2) {};
		\node [style=dot1] (95) at (5, 0) {};
		\node [style=dot1] (96) at (7, 0) {};
		\node [style=dot1] (97) at (5, -2) {};
		\node [style=dot1] (98) at (7, -2) {};
		\node [style=dot1] (99) at (5, -4) {};
		\node [style=dot1] (100) at (7, -4) {};
		\node [style=none] (101) at (4, 0) {$\times$};
		\node [style=none] (102) at (4, -2) {$\times$};
		\node [style=none] (103) at (8.5, 4) {$q_3$};
		\node [style=none] (104) at (8.5, 2) {$q_1$};
		\node [style=none] (105) at (8.5, 0) {$q_2$};
		\node [style=none] (106) at (8.5, -2) {$q_4$};
		\node [style=none] (107) at (8.5, -4) {$q_0$};
	\end{pgfonlayer}
	\begin{pgfonlayer}{edgelayer}
		\draw [style=wire] (82.center) to (83.center);
		\draw [style=wire] (84.center) to (85.center);
		\draw [style=wire] (91.center) to (92.center);
		\draw [style=wire] (2.center) to (5.center);
		\draw [style=wire] (1.center) to (6.center);
		\draw [style=wire] (0.center) to (7.center);
		\draw [style=wire] (3.center) to (70.center);
		\draw [style=wire] (4.center) to (9.center);
		\draw [style=wire] (101.center) to (102.center);
	\end{pgfonlayer}
\end{tikzpicture}$}
    \caption{The final circuit corresponding to the input circuit in Figure \ref{fig:comm-blocks}.}
	\label{fig:comm-blocks-braid-swap}
\end{figure}
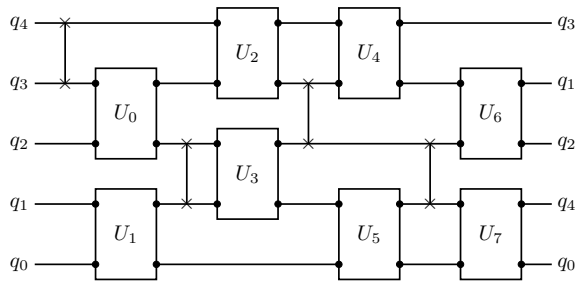

\subsection{The optimization problem: formulation}
\label{section:optim-prob}
We start by defining the constructor generating the permutations $\timeent{P}$ at each time step $t$.
We recall that we denote by $T$ the number of layers of the input circuit, so the layer index $t$ belongs to $\intset{0}{T-1}$.
Also each layer determines $S$ continuous parameters, so overall we have $S\times T$ parameters,
which will be denoted by $\vectheta \in \mathbb{R}^{ST}$.

The constructor configuration (hyper-parameters) is a pair $(p_1, p_2)$ of functions $p_1, p_2: \intset{0}{S - 1} \to V_{\mathcal{M}}$ mapping a sequence index $s$
to a physical qubit\footnote{We also require that $i=p_1(s)\ne p_2(s)=j$ for all $s$, so we always have distinct targets $\{i, j\}$ for the swap.}.
Moreover, in the scope of this section we consider the functions $(p_1, p_2)$ as arbitrary,
however Section \ref{section:line-topo} will develop a specific structure for them.
At each time step $t$ the hyper-parameters determine a sequence of $S$ swaps $\mathrm{SSWAP}_m(p_1(s), p_2(s), \timeent{\theta_s})$
for $s \in \intset{0}{S - 1}$.
So, given a vector of continuous parameters $\vectheta{}=\left(\timeent{\theta_s}\right)_{t, s} \in \mathbb{R}^{ST}$,
the hardware cost function assumes the form
\begin{align}
	\label{eq:circcostfun:psw}
	\mathcal{L}(\boldsymbol{\theta}) =& \sum_{t=0}^{T-1} \mathcal{L}_t(\boldsymbol{\theta}),
\end{align}
with
\begin{subequations}
\begin{align}
	\label{eq:circcostfun-t-term}
	\mathcal{L}_t(\boldsymbol{\theta}) =& \beta(t) \vecop(M_c)^\top \timeent{K} \vecop(\timeent{G}),
\end{align}
and
\begin{align}
	\label{eq:pdecomp}
	\timeent{P} =& \mathrm{PSSWAP}_m(p_1(0), p_2(0), \timeent{\theta_0}) \cdots \mathrm{PSSWAP}_m(p_1(S - 1), p_2(S - 1), \timeent{\theta_{S - 1}}),\\
	\timeent[0]{K} =& \timeent[0]{P},\\
	\timeent{K} =& \timeent{P} \timeent[t-1]{K},\quad 1 \le t \le T - 1\,.
\end{align}
\end{subequations}
In \eqref{eq:circcostfun:psw}, each term is scaled by the function
$\beta: \mathbb{R} \to \mathbb{R}$ which is assumed decreasing and positive valued.
The function $\beta$, is one of the key components of the heuristic solver ---
the \textit{adaptive feasibility}, which is introduced in Section \ref{section:solver}.
Note that each continuous parameter $\timeent[t_1]{\theta_s}$ appears as $\mathrm{PSSWAP}$ argument,
one time for all terms in \eqref{eq:circcostfun:psw} such that $t \ge t_1$.
Also $\mathcal{L}(\boldsymbol{\theta}) \ge 0$ for all $\boldsymbol{\theta}$.

We obtain the ideal form for the optimization problem, that is
\begin{align}
	\label{eq:optim-prob-ideal}
	\underset{\boldsymbol{\theta} \in \mathbb{R}^{ST}}{\mathrm{min}}\quad& \cardf(\boldsymbol{\theta}),\\
	\nonumber
	\mathrm{s.t.}\quad& \mathcal{L}(\boldsymbol{\theta}) = 0,
\end{align}
where $\cardf(\boldsymbol{\theta})$ is the cardinality\footnote{
	The cardinality of a vector $\mathbf{x} \in \mathbb{R}^n$ is defined as
	$\cardf(\mathbf{x})\coloneqq\left|\left\{i\middle| x_i\ne 0\right\}\right|$,
	where $|\cdot|$ is the set cardinality.
}
of the vector $\boldsymbol{\theta}$.
The structure of the problem is justified by the fact that for $\timeent{\theta_s}=0$, the corresponding $\mathrm{PSSWAP}$ is the identity permutation.
The next proposition shows that an optimization problem with cardinality as objective and whose constraint
matches a certain structure, can be solved by an equivalent differentiable problem.
\begin{theorem}
	\label{thm:equiv-optim-prob}
	Given the structure of the hardware cost function $\mathcal{L}$, the following optimization problems are equivalent
	\begin{equation}
		\begin{cases}
			\underset{\vectheta{} \in \mathbb{R}^{ST}}{\mathrm{min}} & \|\vectheta{}\|^2_2,\\
			\text{s.t.}\quad & \mathcal{L}(\vectheta{}) = 0,
		\end{cases}
		\quad \cong \quad
		\begin{cases}
			\underset{\vectheta{} \in \mathbb{R}^{ST}}{\mathrm{min}} & \mathbf{card}(\vectheta{}),\\
			\text{s.t.}\quad & \mathcal{L}(\vectheta{}) = 0\,.
		\end{cases}
	\end{equation}
\end{theorem}
\prooflater{}
This proposition demonstrates that the problem in \eqref{eq:optim-prob-ideal} can be solved by considering the squared $l_2$-norm of $\vectheta$ instead of its cardinality.

Let us study the characterization of the stationary points of the
hardware cost function $\mathcal{L}$. The next proposition shows that vectors $\boldsymbol{\theta} \in \mathbb{R}^{ST}$ such that
the individual elements are integer multiples of $\frac{\pi}{2}$, are a sufficient condition for the stationary points
of the hardware cost function.
\begin{theorem}
	\label{lemma:ell-critical-points}
	Consider the hardware cost function \eqref{eq:circcostfun:psw} and the element $\timeent{\theta_s}$
	with index $(s, t)$ belonging to the vector $\boldsymbol{\theta} \in \mathbb{R}^{ST}$.
	Let $\Omega=\left\{k\frac{\pi}{2}\middle| k \in \mathbb{Z}\right\}$ be the set of integer multiples of $\pi/2$.
	Then
	\begin{itemize}
		\item [1.]
		$\boldsymbol{\theta}^\top \timeent{\mathbf{e}_s} \in \Omega \implies
		\pdv{\mathcal{L}(\boldsymbol{\theta})}{\timeent{\theta_s}} = 0$.
	\end{itemize}
	If $\pdv{\mathcal{L}(\overline{\boldsymbol{\theta}})}{\timeent{\theta_s}} = 0$, for some
	$\overline{\boldsymbol{\theta}} \in \mathbb{R}^{ST}$, then, either
	\begin{itemize}
		\item[2.] $\overline{\boldsymbol{\theta}}^\top \timeent{\mathbf{e}_s} \in \Omega$,
		\item[3.] or $\overline{\boldsymbol{\theta}}^\top \timeent{\mathbf{e}_s} \notin \Omega$
		and
		$\pdv{\mathcal{L}(\widehat{\boldsymbol{\theta}})}{\timeent{\theta_s}} = 0$ for
		$\widehat{\boldsymbol{\theta}} \in \left\{\overline{\boldsymbol{\theta}} + r\timeent{\mathbf{e}_s} \middle| r \in \mathbb{R}\right\}$.
	\end{itemize}
\end{theorem}
\prooflater{}

\begin{remark}
	Note that Case 3 of Proposition \ref{lemma:ell-critical-points} proposition suggests that the constraint is locally ``flat'' in the direction $\mathbf{e}_s$ which means that multiple solutions are possible, but this ambiguity is resolved by $2$-norm minimization which simply picks $0$ in this case. Let us further illustrate Case 3:
	given circuits $C_1$ and $C_2$, we define the equivalence relation $C_1 \sim C_2$ whenever the circuits
	implement the same unitary up to a permutation. Now, assume the line connectivity and consider the swap mapping process that follows.
	\begin{equation}
		\vcenter{\Qcircuit @C=1.0em @R=1.0em @!R {
			& \lstick{q_0:} & \ctrl{2} & \qw\\
			& \lstick{q_1:} & \qw	   & \qw\\
			& \lstick{q_2:} & \targ	   & \qw
		}} \qquad \overset{\mathrm{SWAP}}{\to} \qquad
		\vcenter{\Qcircuit @C=1.0em @R=1.0em @!R {
			& \lstick{q_0:} & \qswap      & \dstick{\scriptstyle\theta_0} \qw & \qw      & \qw & \rstick{q_1} \\
			& \lstick{q_1:} & \qswap \qwx & \qw 						      & \ctrl{1} & \qw & \rstick{q_0} \\
			& \lstick{q_2:} & \qw         & \qw 				              & \targ    & \qw & \rstick{q_2}
		}} \qquad \sim \qquad
		\vcenter{\Qcircuit @C=1.0em @R=1.0em @!R {
			& \lstick{q_0:} & \qswap      & \dstick{\scriptstyle\theta_0} \qw & \qw         & \qw 								& \qw         & \qw & \rstick{q_1} \\
			& \lstick{q_1:} & \qswap \qwx & \qw								  & \qswap      & \dstick{\scriptstyle\theta_1} \qw & \targ		  & \qw & \rstick{q_2} \\
			& \lstick{q_2:} & \qw         & \qw 							  & \qswap \qwx & \qw 								& \ctrl{-1}   & \qw & \rstick{q_0}
		}}
	\end{equation}
	With circuit swap gates interpreted as $\mathrm{SSWAP}$s controlled by a parameter $\theta_k \in \mathbb{R}$.
	The central circuit reports a solution with $\theta_0=\pi/2$,
	however adding another swap ($q_1 \to q_2$) as depicted on the right-hand side
	does not alter the hardware cost function,
	consequently the partial derivative corresponding to the second swap is zero for any $\theta_1 \in \mathbb{R}$.

	In other words, there may exist configurations of the parameters vector $\vectheta$ such that one or more elements of such vector
	determine a flat cost. However, the objective of the optimization problem \eqref{eq:optim-prob-ideal}, being a cardinality,
	favors the zero value for the free parameters.

	\myendofremark{}
\end{remark}

\subsection{The optimization problem: constraint specification for different topologies}
\label{section:line-topo}
In the construction of the hardware cost function $\mathcal L$, we claimed that the sequence of $\mathrm{PSSWAP}$s in \eqref{eq:pdecomp} depends on some hyper-parameters.
In the present section we determine the structure of the $\mathrm{PSSWAP}$s for the case of the \underline{line connectivity} between qubits.
Despite the simplicity of the present topology, the results obtained here are key for the generalization to the arbitrary connectivity.
In relation to general topologies, we notice that recently
there have been a widespread adoption of quantum processing unit topologies based on hexagonal lattices \cite{HeavyHexIBM}.
An example is depicted in Figure \ref{fig:heavy-hex-lattice}. Moreover, the results obtained in this section are extend to the general case
in Appendix \ref{section:arbitrary-topo}.
\begin{figure}
    \centering
	$\begin{tikzpicture}
    [semithick, mydot/.style={inner sep=1pt,circle,draw,fill=gray!30,label={}, minimum size=1.5mm}]
    \newdimen\R
    \R=2cm
    \draw (330:\R) \foreach \x in {30, 90, 150, 210, 270, 330} {  -- node[mydot] [midway] {} (\x:\R) };
    \coordinate (p1) at (240:{cos(30) * 2 * \R});
    \coordinate (p2) at (-60:{cos(30) * 2 * \R});
    \draw ($(p1)+(90:\R)$) \foreach \x in {150, 210, 270, 330, 30} {  -- node[mydot] [midway] {} ($(p1)+(\x:\R)$) };
    \draw ($(p2)+(210:\R)$) \foreach \x in {270, 330, 30, 90} {  -- node[mydot] [midway] {} ($(p2)+(\x:\R)$) };

    \foreach \x in {30, 90, 150, 210, 270, 330} \node[mydot] at (\x:\R) {};
    \foreach \x in {150, 210, 270, 330} \node[mydot] at ($(p1)+(\x:\R)$) {};
    \foreach \x in {270, 330, 30} \node[mydot] at ($(p2)+(\x:\R)$) {};
\end{tikzpicture}$
    \caption{IBM's heavy-hex lattice topology. The graph depicts three unit cells, with vertices representing the qubits,
	and edges the connectivity constraints.}
	\label{fig:heavy-hex-lattice}
\end{figure}
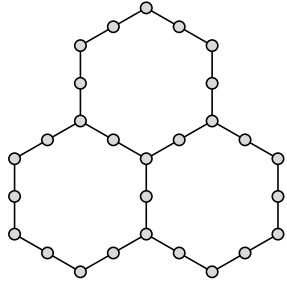

The line connectivity model is defined as a chain of $m$ qubits (assuming $m\ge 3$) where the neighborhoods of qubit $k \in [1\twodots{}m-2]$
are qubits $k-1$ and $k+1$.
The extremes of the chain, that is qubits $0$ and $m-1$, have neighborhood, respectively, qubit $1$ and qubit $m-2$.

Assume for simplicity that the number of qubits $m$ is an odd integer greater than two.
Then the set of generating swaps $\mathcal{P}_{\mathcal{M}}$ defined in \eqref{eq:generating-swaps-set}, contains $m-1$ elements.
We partition the set $\mathcal{P}_{\mathcal{M}}$ into the following subsets
\begin{align}
	\label{eq:line-conn-disjoint-part}
	\Gamma_1=&\left\{ \mathrm{SWAP}(2k, 2k+1) \middle| k \in \left[0\twodots{}(m-1)/2\right]\right\},\\
	\nonumber
	\Gamma_2=&\left\{ \mathrm{SWAP}(2k+1, 2k+2) \middle| k \in \left[0\twodots{}(m-1)/2\right]\right\},
\end{align}
so $\mathcal{P}_{\mathcal{M}} = \Gamma_1 \cup \Gamma_2$.
We note that for any $Q_1, Q_2 \in \Gamma_1$, then $Q_1 Q_2 = Q_2 Q_1$, similarly it also holds for $\Gamma_2$.

The next lemma extends the commutativity from the generating permutations to the generated DSM. The result is immediate,
so we state the claim without proof.
\begin{lemma}
	Let $P_1, P_2$ be $m\times m$ permutation matrices such that $P_1P_2=P_2P_1$, that is the permutations commute.
	Then the following doubly stochastic matrices\footnote{
		Equivalent to the $\mathrm{SSWAP}$ defined in \eqref{eq:sswap-def-cos-sin},
		or the $\mathrm{PSSWAP}$ defined in \eqref{eq:psswap-def-cos-sin}.
	} also commute
	\begin{align}
		Q_k = (1-\alpha_k) \idenm{m} + \alpha_k P_k,
	\end{align}
	with $\alpha_k \in [0, 1]$, for $k=1, 2$. That is, $Q_1Q_2=Q_2Q_1$.
\end{lemma}
Consequently, the product of doubly stochastic matrices obtained from either $\Gamma_1$ or $\Gamma_2$ commute.
Using \eqref{eq:psswapdef} we define the composition of such commuting matrices ($\mathrm{PSSWAP}$s) controlled by the continuous parameters $\boldsymbol{\theta}$, so
\begin{align}
	\label{eq:pswaps-layer-line}
	C(\Gamma_k, \boldsymbol{\theta}) =& \prod_{P_i \in \Gamma_k}
	\left(\idenm{m}^{\otimes 2} + \sin^2(\theta_i) \left(P_i^{\otimes 2} - \idenm{m}^{\otimes 2}\right)\right)\,.
\end{align}
To complete the construction,
we specialize the definition of $\timeent{P}$ related to the hardware cost function in \eqref{eq:pdecomp},
with a (finite) sequence of alternating structures of the form $C(\Gamma_k, \boldsymbol{\theta})$, that is
\begin{align}
	\timeent{P} =&
	C\left(\Gamma_1, \timeent{\boldsymbol{\theta}}_1\right) \cdot
	C\left(\Gamma_2, \timeent{\boldsymbol{\theta}}_2\right) \cdot
	C\left(\Gamma_1, \timeent{\boldsymbol{\theta}}_3\right) \cdot
	C\left(\Gamma_2, \timeent{\boldsymbol{\theta}}_4\right)
	\cdots
\end{align}

\begin{figure}[h]
    \centering
    $\Qcircuit @C=1.0em @R=1.0em @!R {
        & \lstick{q_0:} & \qswap      & \qw &\qw         & \qw\\
        & \lstick{q_1:} & \qswap \qwx & \qw &\qswap      & \qw\\
        & \lstick{q_2:} & \qswap      & \qw &\qswap \qwx & \qw\\
        & \lstick{q_3:} & \qswap \qwx & \qw &\qswap      & \qw\\
        & \lstick{q_4:} & \qw         & \qw &\qswap \qwx & \qw
		\gategroup{1}{3}{5}{3}{1em}{--}
		\gategroup{1}{5}{5}{5}{1.em}{--}
    }$
    \caption{Example of the pattern of swaps for the line connectivity with $m=5$ qubits. The dashed frames represent
	respectively the set $\Gamma_1$ and $\Gamma_2$.}
	\label{fig:line-conn}
\end{figure}
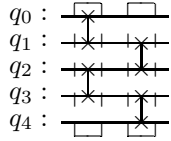

The next lemma shows one of the motivations that justify this construction, that is efficient matrix multiplication for
permutations within the same partition.
\begin{lemma}
	\label{lemma:commuting-perm-mul-as-sum}
	Let $a, b \in \symmg{m}$ be distinct, involutory and commuting elements of the symmetric group of degree $m$,
	that is $a\ne b$, $a^2=b^2=e$ and $a\circ b = b\circ a$, where $e \in \symmg{m}$ is the identity element.
	Let $P_a, P_b$ be the permutation representations \cite{Sagan2001} of $a$ and $b$, respectively.
	Then
	\begin{align}
		\label{eq:commuting-perm-mul-as-sum}
		P_a P_b =& P_a + P_b - \idenm{m}.
	\end{align}
\end{lemma}
\prooflater{}
As a corollary, it can be readily proven that given a (finite) non-empty set of $n$, $m\times m$ permutation matrices $\left\{P_t\right\}_{t=1}^n$,
in which each pair of elements $P_a, P_b$, fulfills the conditions of the Lemma, then
\begin{align}
	\prod_{t=1}^n P_t =& \left(\sum_{t=1}^n P_t\right) - (n - 1)\idenm{m}\,.
\end{align}
We also note that the maximum number of disjoint and thus commuting permutation matrices is $n=\lfloor m/2 \rfloor$.

The next step is that of applying the latest result to a product of doubly stochastic matrices whose generating permutations fulfill
the conditions of Lemma \ref{lemma:commuting-perm-mul-as-sum}.
\begin{theorem}
	\label{thm:dsm-comp-gen-commuting-perm-as-sum}
	Let $\left\{P_t\right\}_{t=1}^n$ be a non-empty set of $m\times m$ matrices such that each element is involutory and each pair commutes.
	Then the following identity holds for the composition of elementary DSM generated by the $P_t$,
	\begin{align}
		\label{eq:dsm-comp-gen-commuting-perm-as-sum}
		\prod_{t=1}^n
		\left(\idenm{m} + \sin^2(\theta_t) \left(P_t - \idenm{m}\right)\right)
		=& \idenm{m} + \sum_{t=1}^n \sin^2(\theta_t)\left(P_t - \idenm{m}\right)\,.
	\end{align}
\end{theorem}
\prooflater{}
Finally we apply the results just obtained to the definition of layer of $\mathrm{PSSWAP}$s \eqref{eq:pswaps-layer-line}, so
\begin{align}
	C(\Gamma_k, \boldsymbol{\theta}) =& \idenm{m}^{\otimes 2} +
	\sum_{P_i \in \Gamma_k}
	\sin^2(\theta_i) \left(P_i^{\otimes 2} - \idenm{m}^{\otimes 2}\right)\,.
\end{align}

Assume we have $m$ qubits, then it can be shown that the repetition of patterns constructed as depicted in Figure \ref{fig:line-conn},
generates a set of permutations that contains a subgroup of the symmetric group $\symmg{m}$.
Also as one may expect, increasing the number of replica creates subgroups that approach in term of order, the group $\symmg{m}$.
So, it would be interesting obtaining results regarding the required number of repetitions and their efficiency.
For now however, we do not expand the latter point which will be addressed in future research.
To conclude, we claim without proving that this construction favors the minimization of the resulting circuit depth.
However, evidence for the latter assertion emerges in the experiments section.

\subsection{The optimization problem: numerical method}
\label{section:solver}
In this section we develop a heuristic for the solution of problem \eqref{eq:optim-prob-ideal}.
We start from a well-known technique called Rolling Horizon (RH).
The latter consists of partitioning a decision problem into a sequence of sub-problems whose aggregated solutions
constitute a solution for the whole problem. The sub-problems often are identified by time windows of fixed length.
Examples of the aforementioned strategy can be found in \cite{rhref1,rhref2}.

We build on the RH strategy to obtain a process we call \textit{adaptive feasibility}.
The terminology finds the following motivation -- we call it adaptive because the RH depth adjusts to the
best sub-problem where we can reach feasibility.
Considering the structure of the hardware cost function, we note that for all $s$, the variables $\timeent[\tau]{\theta_s}$,
appear in the terms $\mathcal{L}_t$ of \eqref{eq:circcostfun:psw} with $t\ge \tau$.
Since the permutation at time $t$ influences the permutations of the subsequent layers, then we favor the feasibility of lower
layers (w.r.t. time $t$) using the decreasing function $\beta(t)$ introduced in \eqref{eq:circcostfun-t-term}.
Now, it follows from Proposition \ref{thm:equiv-optim-prob} that the optimization problem takes the equivalent form
\begin{align*}
	\underset{\vectheta{} \in \mathbb{R}^{ST}}{\mathrm{min}} & \left\|\vectheta{}\right\|^2_2,\\
	\text{s.t.}\quad & \mathcal{L}(\vectheta{}) = 0,
\end{align*}
which we solve using the \textit{Differential Multiplier Method} \cite{NIPS1987_a87ff679}.
The method, by introducing the Lagrange multiplier $\lambda$, produces a sequence of updates for the variables $\vectheta{}$ and $\lambda$, so
\begin{subequations}
\begin{align}
	\vectheta{} \leftarrow& \vectheta{} -\eta_\theta \nabla_\theta \left(\left\|\vectheta{}\right\|^2_2 + \lambda\mathcal{L}(\vectheta{})\right),\\
	\lambda \leftarrow& \lambda +\eta_\lambda \nabla_\lambda \left(\left\|\vectheta{}\right\|^2_2 + \lambda\mathcal{L}(\vectheta{})\right),
\end{align}
\end{subequations}
until a stop condition is reached.
We denoted with $\eta_\theta$ and $\eta_\lambda$ the step sizes for the variables $\vectheta{}$ and $\lambda$, respectively.
Interestingly, since $\mathcal{L}(\vectheta{}) \ge 0$, the second update corresponds to a monotonic increase of $\lambda$, that is
\begin{align}
	\lambda \leftarrow& \lambda +\eta_\lambda \mathcal{L}(\vectheta{})\,.
\end{align}
The latter can be used to prove that the optimizer gets attracted by the stationary points of $\mathcal{L}$, but according to
Proposition \ref{lemma:ell-critical-points}, such points have a known and convenient structure.

The entire procedure is split into two algorithms, the \texttt{DSM-SWAP} and the \texttt{Knitter}, with the former being the main algorithm.

\subsubsection{The \texttt{Knitter} algorithm}
The \texttt{Knitter}\footnote{The name \texttt{Knitter} is inspired by the braid diagrams (Figure \ref{fig:braids-horizon})
used to represent the iteration of the SWAPs.} algorithm can be interpreted as a global solver (w.r.t. the circuit) for the swap mapping problem.
But the input circuit is split into sections using the RH strategy so the \texttt{Knitter} only acts upon each of the sub-circuits separately.

We describe the steps of the algorithm as presented in Algorithm \ref{algo:knitter}.
The input circuit is given as a sequence of graphs $\timeent{\mathcal{G}}$, the latter is used alongside the hardware graph $\mathcal{M}$
to construct (function $\mathtt{BuildHardwareCost}$) the hardware cost function $\mathcal{L}$.
The construction depends on the machine topology, details are elaborated in sections \ref{section:arbitrary-topo} and \ref{section:line-topo}.

In the for-loop at line \ref{alg:knitter:optim-loop} the vector $\vectheta{}$ is updated in a gradient descent fashion using the gradient of the
function $f(\vectheta{})=\left\|\vectheta{}\right\|^2_2 + \lambda\mathcal{L}(\vectheta{})$.
Since the problem is non-convex, the iteration is executed $\mathtt{max\_trials} \ge 1$ times.
At the end of each iteration,
the \textit{projection}\footnote{The definition of \textit{projection} is given in the proof of Lemma \ref{lemma:min-card}.}
$\projop{\Omega^{ST^\prime}}$ onto the set $\Omega^{ST^\prime}$
is applied to the vector $\vectheta{}$, with $\Omega$ defined as in Proposition \ref{lemma:ell-critical-points}.
Furthermore, the application of the projector is justified by Proposition \ref{lemma:ell-critical-points}.

Completed the trials at line \ref{alg:knitter:end-of-trials},
we use the merit function $g(\vectheta)=\left\|\vectheta{}\right\|^2_2 + \alpha\mathcal{L}(\vectheta{})$ to choose the best solution
\footnote{In our specific implementation we choose the solution that maximizes the number of feasible layers,
consequently the adaptive horizon step is maximized, leading to a reduction of the overall computation time for the method.}.
Here the parameter $\alpha>0$ is a trade-off between swaps minimization and feasibility maximization.
Once the best solution is realized, at line \ref{alg:knitter:count-l} we count the number $l$ of subsequent layers, starting from the first,
that fulfill the hardware constraints.
Here we denote with $\delta_{\{0\}}$ the indicator function\footnote{The indicator function for a subset $C \subseteq U$ is defined as
the extended real-valued function $\delta_C: U \to \mathbb{R} \cup \{\infty\}$ with rule
$\delta_C(\mathbf{x})=\begin{cases}0,&\mathbf{x}\in C,\\\infty,&\text{otherwise}.\end{cases}$} for the set $\{0\} \subset \mathbb{R}$.
Remarkably, line \ref{alg:knitter:count-l} constitutes one of the key elements for the adaptive feasibility.
Finally the algorithm returns the sub-vector of $\vectheta{}^*$ corresponding to the first $l$ layers.

\begin{algorithm}[h]
	\footnotesize
	\caption{The pseudocode for the \texttt{Knitter} algorithm.}
	\label{algo:knitter}
	\SetAlgoLined
	\KwData{Sequence of $T^\prime$ circuit layers as graphs $\timeent{\mathcal{G}}$.
	Hardware connectivity graph $\mathcal{M}$.}
	\SetKwData{Maxsteps}{max\_optim\_steps}\SetKwData{Samples}{samples}
	\SetKw{Break}{break}
	\SetKw{Nil}{nil}
	\SetKw{Is}{is}
	\KwResult{$\vectheta{}^*$\tcp*{Parameters for the feasible layers}}
	$R \leftarrow \emptyset$\tcp*{Set of solutions from trials}
	$\mathcal{L} \leftarrow \mathtt{BuildHardwareCost}\left(\,\left(\timeent{\mathcal{G}}\right)_{t=1}^{T^\prime}, \mathcal{M}\right)$
	\tcp*{Prepare the hardware cost function}
	\smallskip
	\For{$t \leftarrow 1$ \KwTo $\mathtt{max\_trials}$}{
		$\vectheta{} \sim \mathsf{U}_{ST^\prime}\left(0, \epsilon\right)$
		\tcp*{Sample initial $\vectheta{} \in \mathbb{R}^{ST^\prime}$, with $\epsilon>0$ a small constant.}
		$\lambda \sim \mathsf{U}\left(0, \epsilon\right)$
		\tcp*{Sample initial $\lambda$. $\mathsf{U}$ is the uniform distribution.}
		\For{$\tau \leftarrow 1$ \KwTo \Maxsteps}{\label{alg:knitter:optim-loop}
			$\vectheta{} \leftarrow \vectheta{} -\eta_\theta \nabla_\theta \left(\left\|\vectheta{}\right\|^2_2 + \lambda\mathcal{L}(\vectheta{})\right)$
			\tcp*{Update $\mathbf{\theta}$ with GD step}
			$\lambda \leftarrow \lambda +\eta_\lambda \mathcal{L}(\vectheta{})$
			\tcp*{Update $\lambda$}
			\If{$\left\|\nabla_\theta \mathcal{L}(\vectheta{})\right\|^2_2 \le \gamma$}{\Break\tcp*{Early stopping condition met}}
		}
		$R \leftarrow R \cup \left\{\projop{\Omega^{ST^\prime}}(\vectheta{})\right\}$
		\tcp*{Store projection of current $\vectheta{}$}
	}
	$\vectheta{}^* \leftarrow \underset{\vectheta{} \in R}{\mathrm{argmin}}\, \left\|\vectheta{}\right\|^2_2 + \alpha\mathcal{L}(\vectheta{})$
	\label{alg:knitter:end-of-trials}
	\tcp*{Best solution selection policy}
	$l \leftarrow \underset{k\in \intset{0}{T^\prime}}{\mathrm{argmin}}\,\delta_{\{0\}}\left(\sum_{i=0}^{k-1}\mathcal{L}_i(\vectheta{}^*)\right) - k$
	\label{alg:knitter:count-l}
	\tcp*{Count feasible layers in $l$}
	\If{$l$ \Is $0$}{
		\tcp{No feasible layers}
		$\vectheta{}^* \leftarrow $\Nil\;
		\Return{}\;
	}
	$\vectheta{}^* \leftarrow \vectheta{}^*\left[1\isep l*S\right]$
	\tcp*{Select parameters (sub-vector) for the $l$ feasible layers}
\end{algorithm}

\subsubsection{The \texttt{DSM-SWAP} algorithm}
The \texttt{DSM-SWAP} algorithm in essence partitions the input circuit into sub-circuits upon which the \texttt{Knitter} algorithm
is executed. The pseudocode is presented in Algorithm \ref{algo:dsm-swap}, also we recall that we denote with $T$
the number of layers of the circuit and with $S$ the number of parameters (equivalently $\mathrm{SWAP}$s) per layer.
The depth of the sub-circuits is given by the hyper-parameter $\mathtt{horizon}$, however
the starting point for the horizon advances adaptively (line \ref{alg:dsm-swap:update-hori-start}) depending on the feasibility reached
by the previous iteration.

Function $\mathtt{ThetasToSwaps}$ invoked at lines \ref{alg:dsm-swap:current-theta-to-swaps} and \ref{alg:dsm-swap:final-theta-to-swaps}
takes a vector of angles $\vectheta{}$ to a sequence of permutations matrices. We note that the elements of the vector $\vectheta{}$
are expected to belong to the set $\Omega$ (Proposition \ref{lemma:ell-critical-points}), that is the angles represent a vertex of the
Birkhoff polytope. But this is consistent with the value returned by function $\mathtt{Knitter}$.

We remark one additional point at line \ref{alg:dsm-swap:bullet-action}. The permutations applied to layer $t$, influence all the subsequent
layers from $t+1$ to $T-1$. Consequently, we make the algorithm consistent with the mechanism by pre-permuting the qubits of each block $C$
with the permutations from the previous layers.
In the specific expression at line \ref{alg:dsm-swap:bullet-action} we denote with $(\cdot)\,\bullet\,(\cdot)$
the action of the permutation $P$ on the circuit $C$.

Finally the result of the method consists of a sequence of $T\times S$ involutory permutation matrices (either identity or $\mathrm{SWAP}$).
In Figure \ref{fig:heavy-hex-lattice} the reader can appreciate the resulting structure visually.

\begin{algorithm}
	\footnotesize
	\caption{The pseudocode for the DSM-SWAP algorithm.}
	\label{algo:dsm-swap}
	\SetAlgoLined
	\KwData{Sequence of $T$ circuit layers as graphs $\timeent{\mathcal{G}}$.
	Hardware connectivity graph $\mathcal{M}$.}
	\SetKwData{Samples}{samples}
	\SetKw{Nil}{nil}
	\SetKw{Is}{is}
	\KwResult{$R$\tcp*{Sequences of swaps for each layer}}
	$\vectheta{}^* \leftarrow (\,)$
	\tcp*{Init empty vector ($\mathrm{dim}\,\vectheta{}^*=0$) for the parameters}
	$t \leftarrow 0$\;
	\smallskip
	\While{$t<T$}{
		\tcp{Iterate over circuit layers}
		$h \leftarrow \min\left(\mathtt{horizon}, T - t\right)$
		\tcp*{Compute effective horizon}
		$C \leftarrow \left(\timeent[\tau]{\mathcal{G}}\right)_{\tau=t}^{t+h}$
		\tcp*{Construct sub-circuit with depth up to $\mathtt{horizon}$}
		$P \leftarrow \mathtt{ThetasToSwaps}(\vectheta{}^*)$\label{alg:dsm-swap:current-theta-to-swaps}\;
		$C \leftarrow P \bullet C$
		\label{alg:dsm-swap:bullet-action}
		\tcp*{Apply the resulting permutation up to layer $t$}
		$\vectheta{} \leftarrow \mathtt{Knitter}\left(C, \mathcal{M}\right)$
		\tcp*{Run $\mathtt{Knitter}$ on permuted sub-circuit $C$}
		\If{$\vectheta{}$ \Is \Nil}{
			\tcp{If there are no feasible layers}
			\tcp{fail or run alternative strategy}
			$R \leftarrow $ \Nil\tcp*{Nil result}
			\Return{}\;
		}
		$t \leftarrow t + (\mathrm{dim}\,\vectheta{})/S$
		\label{alg:dsm-swap:update-hori-start}
		\tcp*{Adaptively update the next horizon starting point}
		$\vectheta{}^* \leftarrow \vectheta{}^* \oplus \vectheta{}$
		\tcp*{Extend the vector of parameters}
		\tcp{Note $\mathrm{dim}\,\vectheta{}^* = t\times S \le T\times S$.}
	}
	\tcp{The final $\vectheta{}^*$ belongs to $\mathbb{R}^{ST}$}
	$R \leftarrow \mathtt{ThetasToSwaps}(\vectheta{}^*)$
	\label{alg:dsm-swap:final-theta-to-swaps}
	\tcp*{Obtain swaps for all layers}
\end{algorithm}

\section{Experiments}
\label{section:experiments}
The purpose of the experiments is two fold. On one hand we aim at obtaining a clearer view of the effects of the
hyper-parameters. On the other one, we compare the results of the proposed method with other well known algorithms in
literature.

The algorithm DSM-SWAP has been implemented\footnote{Source code available at \url{https://github.com/qiskit-community/dsm-swap}.}
on top of the frameworks Qiskit \cite{Qiskit} and PyTorch \cite{pytorch}.
From the Qiskit library, we also made use of the algorithm SABRE and the circuit generators for the multi-controlled X gate
and Quantum Volume.

The circuits involved in the experiments may present features with different distributions,
depending on patterns and number of qubits. Also the measurements we are considering are taken both
before and after the application of the swap mapping process.
To evaluate the generalization of the strategy we consider several circuit patterns, specifically
the Quantum Volume (QV\footnote{In the interest of space, we denote with $\mathrm{QV}\{n\}$ a Quantum Volume circuit with $n$ qubits.
Moreover, the latter is generated via the function \texttt{qiskit.circuit.quantumcircuit.QuantumCircuit(n, seed=seed)},
where \texttt{seed} determines the circuit instance.})
circuits \cite{QV} and the multi-cX gates compiled with an alphabet consisting of $\mathrm{CNOT}$ and single qubit gates.
Given a circuit pattern $C$ and a feature $X$, we denote with $X_0(C)$ and $X_s(C)$ the random variables
for the feature $X$ measured on $C$, respectively before and after the swap mapping.
Then given the samples $x_0 \sim X_0(C)$ and  $x_s \sim X_s(C)$, assuming $x_0 \ne 0$,
we define the relative measure as $d=\frac{x_s - x_0}{x_0}$.
The features we consider are the number of $\mathrm{CNOT}$s and
the circuit depth\footnote{We consider the Qiskit \cite{Qiskit} definition of circuit depth which can be
be obtained through the method \texttt{qiskit.circuit.QuantumCircuit.depth}.}.
The features are computed on the circuit resulting from the invocation of the Qiskit \footnote{Qiskit 0.36.1 and Qiskit Terra 0.20.1.}
transpiler with optimization level three\footnote{Specifically, the structure of the invocation is \texttt{qiskit.compiler.transpile(..., optimization\_level=3)}.}.
We note that the transpiler embodies algorithms based on the stochastic approach, thus modeling using random variables is justified.
Take for example the number of $\mathrm{CNOT}$s and let $c_0$, $c_s$ be their count before and after the swap mapping.
Then the relative measure $\mathtt{dcnots}=\frac{c_s - c_0}{c_0}$ represents the fractional increase in the number of $\mathrm{CNOT}$s
as a result of the qubit allocation. So for example, if the circuit prior to the swap mapping contains 100 CNOTs and the process
produces a fractional increase of 0.75 units ($\mathtt{dcnots}=0.75$), then the final circuits contains 175 CNOTs.
Similarly we denote the fractional increase in depth with $\mathtt{ddepth}$.

\subsection{Study of the hyper-parameters}
We start by describing the structure of the sampling.
We generate 250 QV circuit instances for each qubit count from 5 to 8 (both inclusive).
Each circuit instance is then processed by the DSM-SWAP algorithm configured with combinations
of increasing horizon $\{1, 2, 4\}$ and maximum optimizer steps $\{10, 30, 100\}$.

In Figure \ref{fig:qv-cmp-hori} we highlight the general positive effect of a longer horizon on both \texttt{dcnots} and \texttt{ddepth}.
Moreover, Figure \ref{fig:braids-horizon} provides the same evidence using braid diagrams.
In Figure \ref{fig:qv-cmp-max-optim-steps} we observe vague evidence that a higher number of steps increases the quality of the results.
This means that the optimizer converges quite fast so we conclude that there is not much difference between 30 and 100 steps, thus the former
value is set as the default.

\subsection{DSM-SWAP vs SABRE}
\begin{figure}
	\centering
	\scalebox{0.70}{\input{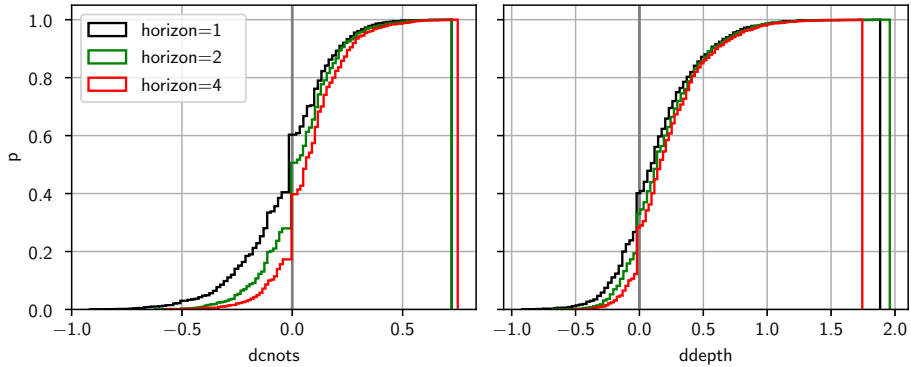}}
	\caption[Experiments]{
		The eCDFs for features gap $\mathtt{dcnots}$ (left) and $\mathtt{ddepth}$ (right) for the comparative case DSM-SWAP vs SABRE.
		The DSM-SWAP algorithm is configured with increasing horizon ($\mathtt{horizon}=1, 2, 4$) and $\mathtt{max\_optim\_steps=30}$.
		The vertical gray line at abscissa zero determines the threshold where the two algorithms produce the same result (w.r.t. the current feature),
		whereas on the right-hand side, the DSM-SWAP yields more favorable results.
	}
	\label{fig:qv_all-cmp-hori-sabre}
\end{figure}

\begin{figure}
	\centering
	\begin{subfigure}[b]{\textwidth}
		\centering
		\scalebox{0.70}{\input{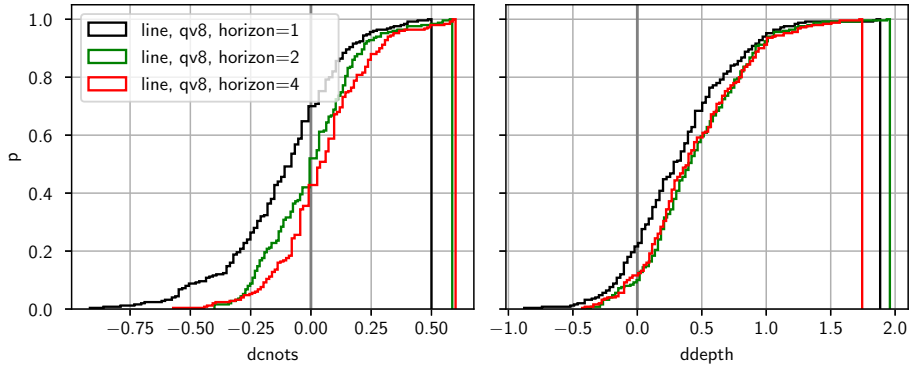}}
		\caption{QV8-line}
		\label{fig:qv8-line-cmp-hori-sabre}
	\end{subfigure}
	\begin{subfigure}[b]{\textwidth}
		\centering
		\scalebox{0.70}{\input{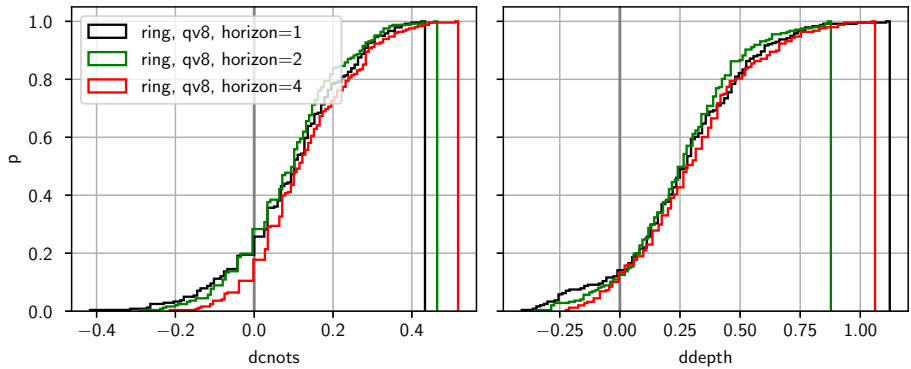}}
		\caption{QV8-ring}
		\label{fig:qv8-ring-cmp-hori-sabre}
	\end{subfigure}
	\caption[Experiments]{
		The same experiment as that in Figure \ref{fig:qv_all-cmp-hori-sabre} except that here we distinguish the case with
		QV8 circuits with line (top row) and ring (bottom row) coupling maps.
	}
	\label{fig:qv8-cmp-hori-sabre}
\end{figure}

We compare the new method with the algorithm SABRE configured with the lookahead strategy \cite{sabre}.
The circuits considered are the Quantum Volume and the multi-cX gates.
The data for DSM-SWAP is the same one obtained for the hyper-parameters investigations\footnote{We fix the hyper-parameter $\mathtt{max\_optim\_steps}=30$.}
we denote the corresponding features with $\mathtt{dcnots}_{dsm}$ and $\mathtt{ddepth}_{dsm}$.
In addition we execute SABRE on the
same circuits and measure the features $\mathtt{dcnots}_{sabre}$ and $\mathtt{ddepth}_{sabre}$.
In Figures \ref{fig:qv_all-cmp-hori-sabre} and \ref{fig:qv8-cmp-hori-sabre}, we plot the eCDFs for the features gaps
\begin{subequations}
\begin{align}
	\mathtt{dcnots} =& \mathtt{dcnots}_{sabre} - \mathtt{dcnots}_{dsm}\\
	\mathtt{ddepth} =& \mathtt{ddepth}_{sabre} - \mathtt{ddepth}_{dsm}.
\end{align}
\end{subequations}
Since the selected features, the smaller they are the better the performance, then points on the positive abscissa correspond to DSM-SWAP
performing better than SABRE. In Figure \ref{fig:qv_all-cmp-hori-sabre} we see that for $\mathtt{horizon} \ge 2$, the new method has at least $80\%$
chances to produce a shallower circuit. In Figure \ref{fig:qv8-cmp-hori-sabre} we distinguish the results for QV8 circuits assuming line and ring couplings.
Surprisingly, in the ring connectivity case, the gap is remarkable even for $\mathtt{horizon}=1$. We think that the latter could be an important clue for
the development of new swap mapping methods.

\begin{figure}
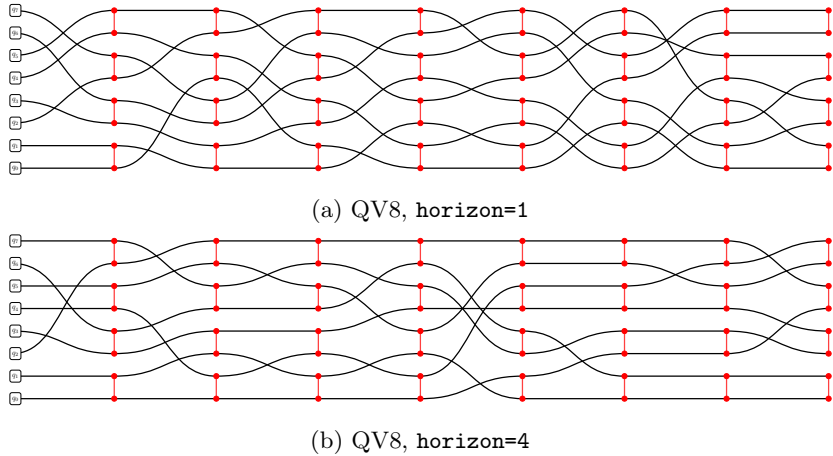

	\centering
	\begin{subfigure}[b]{\textwidth}
		\centering
		\scalebox{0.24}{\input{braid1-h1.pgf}}
		\caption{QV8, \texttt{horizon=1}}
	\end{subfigure}
	\begin{subfigure}[b]{\textwidth}
		\centering
		\scalebox{0.24}{\input{braid1-h4.pgf}}
		\caption{QV8, \texttt{horizon=4}}
	\end{subfigure}
	\caption[Braid diagrams]{\textit{Braid diagrams} for the swap mapping applied to a Quantum Volume circuit with 8 qubits.
	The results, starting from the top, correspond to respectively horizon 1 and 4.
	It can be noticed that in the bottom case the density of the braids (permutations) is visibly lower.}
	\label{fig:braids-horizon}
\end{figure}

\begin{figure}
	\centering
	\scalebox{0.70}{\input{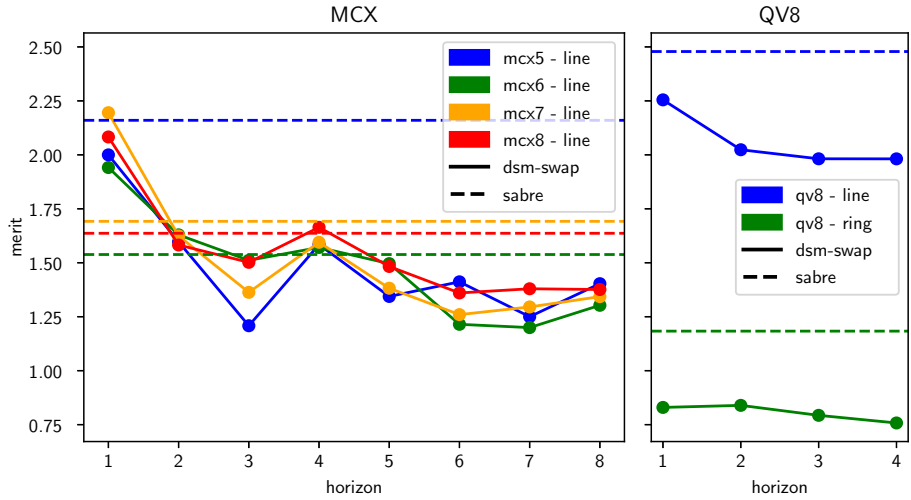}}
	\caption[Experiments]{
		Comparison DSM-SWAP vs SABRE w.r.t. the merit function defined in \eqref{eq:merit-ddcnots-ddepth},
		on MCX (left) and QV8 (right) circuits.
	}
	\label{fig:mcx-qv8-avg-merit}
\end{figure}

In Figure \ref{fig:mcx-qv8-avg-merit} we obtain the average values for the merit measure defined as
\begin{align}
	\label{eq:merit-ddcnots-ddepth}
	\mathtt{merit} =& \mathtt{dcnots} + \mathtt{ddepth},
\end{align}
which can be interpreted as the overall increase in both CNOT count and depth, as a consequence of the swap mapping.
The latter figure shows how our method compares to SABRE as the $\mathtt{horizon}$ increases, also we consider MCX and QV8 circuits.
\begin{remark}
	We expand on the reason for the choice of QV and MCX circuits.
	The Quantum Volume circuits have the property that each layer is represented by a graph $\timeent{\mathcal{G}}$
	that has the maximum number of edges such that the corresponding two-qubit gates commute.
	This feature can be appreciated in Figure \ref{fig:braids-horizon} by observing the columns (layers) of red edges.

	On the other hand, MCX circuits, compiled with an alphabet of CNOT and single qubit gates, present layers that tend to have
	a single edge, consequently they represent the opposing case to QV.

	\myendofremark{}
\end{remark}
Following the previous remark we assert that the case of the QV circuits is the most advantageous for DSM-SWAP since the construction of the
$\mathrm{SWAP}$s (Sections \ref{section:line-topo} and \ref{section:arbitrary-topo}) can maximize the parallelism of the permutations.
The MCX circuits instead should be the opposing case, indeed we observe in Figure \ref{fig:mcx-qv8-avg-merit}
that we need a longer horizon to obtain a neat advantage over SABRE.

\begin{figure}
	\centering
	\begin{subfigure}[b]{\textwidth}
		\centering
		\scalebox{0.70}{\input{qv_all-cmp-hori.pgf}}
		\caption{$\mathrm{QV}\{5, 6, 7, 8\}$}
		\label{fig:qv_all-cmp-hori}
	\end{subfigure}
	\begin{subfigure}[b]{\textwidth}
		\centering
		\scalebox{0.70}{\input{qv8-cmp-hori.pgf}}
		\caption{$\mathrm{QV}8$}
		\label{fig:qv8-cmp-hori}
	\end{subfigure}
	\caption[Experiments]{The eCDFs for features $\mathtt{dcnots}$ (left column) and $\mathtt{ddepth}$ (right column) measured on the algorithm DSM-SWAP.
	The top row configuration consists of
	increasing horizon ($\mathtt{horizon}=1, 2, 4$)
	over $\mathrm{QV}\{5, 6, 7, 8\}$ circuits
	and maximum optimizer steps fixed to 30.
	The bottom row corresponds to the same configuration, however it relates exclusively to the QV8 circuits.}
	\label{fig:qv-cmp-hori}
\end{figure}

\begin{figure}
	\centering
	\begin{subfigure}[b]{\textwidth}
		\centering
		\scalebox{0.70}{\input{qv_all-cmp-optim_steps.pgf}}
		\caption{$\mathrm{QV}\{5, 6, 7, 8\}$}
		\label{fig:qv_all-cmp-optim_steps}
	\end{subfigure}
	\begin{subfigure}[b]{\textwidth}
		\centering
		\scalebox{0.70}{\input{qv8-cmp-optim_steps.pgf}}
		\caption{$\mathrm{QV}8$}
		\label{fig:qv8-cmp-optim_steps}
	\end{subfigure}
	\caption[Experiments]{The eCDFs for features $\mathtt{dcnots}$ (left column) and $\mathtt{ddepth}$ (right column) measured on the algorithm DSM-SWAP.
	The top row configuration consists of
	increasing maximum optimizer steps ($\mathtt{max\_optim\_steps}=10, 30, 100$)
	over $\mathrm{QV}\{5, 6, 7, 8\}$ circuits
	and horizon fixed to 2.
	The bottom row corresponds to the same configuration, however it relates exclusively to the QV8 circuits.}
	\label{fig:qv-cmp-max-optim-steps}
\end{figure}

\begin{figure}
	\centering
	\scalebox{0.7}{\input{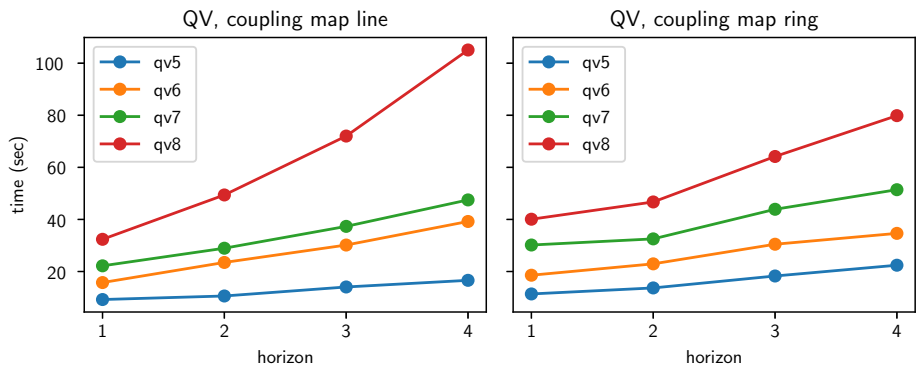}}
	\caption[Experiments]{
		Average time for the DSM-SWAP obtained at increasing horizon with QV circuits.
	}
\end{figure}

\section{Conclusions}
\label{section:conclusions}
The swap mapping process is fundamental for the quantum compiler and increasing its efficiency is essential for
improving the performance of the hardware.
In this work we obtained a procedure that combines both mathematical optimization and heuristic strategies.
In literature however, methods based on mathematical optimization have not recorded particular successes when applied
to this problem. Conscious of the computation complexity hardness of the problem we understand that, to be practical,
the algorithm must include an heuristic component.

The decision process for the insertion of the $\mathrm{SWAP}$s has been modeled as a smooth optimization based on
doubly stochastic matrices.
Also we obtained a procedure to build the pattern of $\mathrm{SWAP}$s depending on the topology of the target hardware.
One of the bottlenecks in the evaluation of the cost function is determined by the number of matrix multiplications,
however, using the commutativity properties of the pattern we devised an efficient scheme for reducing the computational cost.
The solver heuristic is inspired by the rolling horizon policy and it offers linear scaling of the
computational complexity with respect to the depth of the circuit.
The algorithm has been implemented on top of the frameworks Qiskit and PyTorch, also we made the code open source.
The experiments revealed clues regarding the algorithm hyper-parameters and its relation with the state of the art
algorithm SABRE.
The optimality of the results has been measured using the relative increment in $\mathrm{CNOT}$ gate count and depth.
For the hyper-parameters, we found that an increased horizon for the adaptive feasibility influences positively
the performance of the results.
The comparison with the algorithm SABRE shown that at the cost of increased computational time and while preserving the
number of $\mathrm{CNOT}$s, the new method delivers significant reduction in the depth of the resulting circuit.
We see the potential application of our method to the compiling of quantum libraries where the processing time can be
penalized in favor of depth and $\mathrm{CNOT}$s optimality.

Despite the positive results, further research is required to extend the applicability of the method to the upcoming quantum
hardware where we expect the number of qubits to climb to hundreds if not thousands.

\section*{Acknowledgment}
This study has received funding from the Disruptive Technologies Innovation Fund (DTIF),
by Enterprise Ireland, under project number DTIF2019-090 (project QCoIR).
The literature review is based on some unpublished work by
Claudio Gambella, Andrea Simonetto, Anton Dekusar and Giacomo Nannicini.
The open source implementation was realized thanks to the collaboration of
Anton Dekusar, Albert Akhriev and Kevin Krsulich.
The authors are highly thankful to Ali Javadi-Abhari, Dr. Martin Mevissen and Tara Murphy
for their support and suggestions.

\bibliographystyle{plain}
\bibliography{refs}

\begin{thebibliography}{10}

\bibitem{Qiskit}
Qiskit: An open-source framework for quantum computing, 2021.

\bibitem{BeckBook}
Amir Beck.
\newblock {\em First-Order Methods in Optimization}.
\newblock SIAM-Society for Industrial and Applied Mathematics, Philadelphia,
  PA, USA, 2017.

\bibitem{brualdi_2006}
Richard~A. Brualdi.
\newblock {\em Combinatorial Matrix Classes}.
\newblock Encyclopedia of Mathematics and its Applications. Cambridge
  University Press, 2006.

\bibitem{param-shift-rule}
Gavin~E. Crooks.
\newblock Gradients of parameterized quantum gates using the parameter-shift
  rule and gate decomposition.
\newblock 2019.

\bibitem{QV}
Andrew~W. Cross, Lev~S. Bishop, Sarah Sheldon, Paul~D. Nation, and Jay~M.
  Gambetta.
\newblock Validating quantum computers using randomized model circuits.
\newblock {\em Physical Review A}, 100(3), sep 2019.

\bibitem{rhref2}
Claudio Gambella, Enrico Malaguti, Filippo Masini, and Daniele Vigo.
\newblock Optimizing relocation operations in electric car-sharing.
\newblock {\em Omega}, 81:234--245, 2018.

\bibitem{matanalysis}
Roger~A. Horn and Charles~R. Johnson.
\newblock {\em Matrix Analysis}.
\newblock Cambridge University Press, USA, 2nd edition, 2012.

\bibitem{sabre}
Gushu Li, Yufei Ding, and Yuan Xie.
\newblock Tackling the qubit mapping problem for nisq-era quantum devices.
\newblock In {\em Proceedings of the Twenty-Fourth International Conference on
  Architectural Support for Programming Languages and Operating Systems}, pages
  1001--1014, 2019.

\bibitem{Magesan_2012}
Easwar Magesan, Jay~M. Gambetta, and Joseph Emerson.
\newblock Characterizing quantum gates via randomized benchmarking.
\newblock {\em Physical Review A}, 85(4), apr 2012.

\bibitem{maslov-qubit-placement}
Dmitri Maslov, Sean~M Falconer, and Michele Mosca.
\newblock Quantum circuit placement.
\newblock {\em IEEE Transactions on Computer-Aided Design of Integrated
  Circuits and Systems}, 27(4):752--763, 2008.

\bibitem{nannicini2021optimal}
Giacomo Nannicini, Lev~S Bishop, Oktay Gunluk, and Petar Jurcevic.
\newblock Optimal qubit assignment and routing via integer programming, 2021.

\bibitem{rhref1}
Rodrigo Palma-Behnke, Carlos Benavides, Fernando Lanas, Bernardo Severino,
  Lorenzo Reyes, Jacqueline Llanos, and Doris Sáez.
\newblock A microgrid energy management system based on the rolling horizon
  strategy.
\newblock {\em IEEE Transactions on Smart Grid}, 4(2):996--1006, 2013.

\bibitem{pytorch}
Adam Paszke, Sam Gross, Francisco Massa, Adam Lerer, James Bradbury, Gregory
  Chanan, Trevor Killeen, Zeming Lin, Natalia Gimelshein, Luca Antiga, Alban
  Desmaison, Andreas Kopf, Edward Yang, Zachary DeVito, Martin Raison, Alykhan
  Tejani, Sasank Chilamkurthy, Benoit Steiner, Lu~Fang, Junjie Bai, and Soumith
  Chintala.
\newblock Pytorch: An imperative style, high-performance deep learning library.
\newblock In H.~Wallach, H.~Larochelle, A.~Beygelzimer, F.~d\textquotesingle
  Alch\'{e}-Buc, E.~Fox, and R.~Garnett, editors, {\em Advances in Neural
  Information Processing Systems 32}, pages 8024--8035. Curran Associates,
  Inc., 2019.

\bibitem{NIPS1987_a87ff679}
John Platt and Alan Barr.
\newblock Constrained differential optimization.
\newblock In D.~Anderson, editor, {\em Neural Information Processing Systems},
  volume~0. American Institute of Physics, 1987.

\bibitem{HeavyHexIBM}
IBM Quantum.
\newblock The ibm quantum heavy hex lattice.
\newblock Available at \url{https://research.ibm.com/blog/heavy-hex-lattice}
  (2021/07/07).

\bibitem{Sagan2001}
Bruce~E. Sagan.
\newblock {\em The Symmetric Group: Representations, Combinatorial Algorithms,
  and Symmetric Functions}.
\newblock Springer New York, New York, NY, 2001.

\bibitem{qubit-alloc}
Marcos~Yukio Siraichi, Vinicius Fernandes~Dos Santos, Caroline Collange, and
  Fernando~Magno Quint{\~a}o~Pereira.
\newblock {Qubit Allocation}.
\newblock In {\em {CGO 2018 - International Symposium on Code Generation and
  Optimization}}, pages 1--12, Vienna, Austria, February 2018.

\bibitem{Tan_2021}
Bochen Tan and Jason Cong.
\newblock Optimality study of existing quantum computing layout synthesis
  tools.
\newblock {\em {IEEE} Transactions on Computers}, 70(9):1363--1373, sep 2021.

\bibitem{VIZING}
V.~G. Vizing.
\newblock Critical graphs with given chromatic class.
\newblock {\em Metody Diskret. Analiz.}, 5:9--17, 1965.

\end{thebibliography}

\appendix
\section{Appendix}

\subsection{Arbitrary topology}
\label{section:arbitrary-topo}
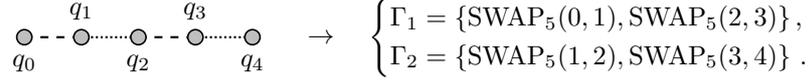
\begin{figure}[h]
    \centering
	$\begin{tikzpicture}
    [mydot/.style={circle, draw, fill=gray!50, inner sep=1pt, semithick, minimum size=2mm}]
    \draw [thick, dashed] (0, 0) -- (1.5, 0);
    \draw [thick, densely dotted] (1.5, 0) -- (3, 0);
    \draw [thick, dashed] (3, 0) -- (4.5, 0);
    \draw [thick, densely dotted] (4.5, 0) -- (6, 0);

    \node[mydot, label=below:$q_0$] at (0, 0) {};
    \node[mydot, label=above:$q_1$] at (1.5, 0) {};
    \node[mydot, label=below:$q_2$] at (3, 0) {};
    \node[mydot, label=above:$q_3$] at (4.5, 0) {};
    \node[mydot, label=below:$q_4$] at (6, 0) {};
\end{tikzpicture}\quad\rightarrow\quad
	\begin{cases}
		\Gamma_1 = \left\{\swapop{5}(0, 1), \swapop{5}(2, 3)\right\},\\
		\Gamma_2 = \left\{\swapop{5}(1, 2), \swapop{5}(3, 4)\right\}\,.
	\end{cases}$
    \caption{An example of edge coloring for the line connectivity.
	On the LHS we have the graph $\mathcal{M}$ with the line pattern representing the color of the edge.
	On the RHS we have the partitions obtained in \eqref{eq:line-conn-disjoint-part}.}
	\label{fig:line-edge-coloring-to-gens}
\end{figure}

We extend the results obtained in the previous section from line connectivity to arbitrary topology.
The disjoint partitioning $\mathcal{P}_{\mathcal{M}} = \Gamma_1 \cup \Gamma_2$ of the set of generating swaps,
obtained in \eqref{eq:line-conn-disjoint-part},
can be interpreted as a special case of the \textit{graph edge coloring problem}.
In Figure \ref{fig:line-edge-coloring-to-gens} we provide an example of line connectivity related to the partitioning and its relation
to edge coloring.

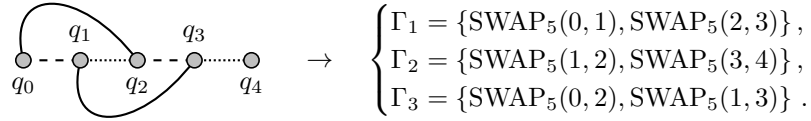
\begin{figure}[h]
    \centering
	$\begin{tikzpicture}
    [mydot/.style={circle, draw, fill=gray!50, inner sep=1pt, semithick, minimum size=2mm}]
    \draw [thick, dashed] (0, 0) -- (1.5, 0);
    \draw [thick, densely dotted] (1.5, 0) -- (3, 0);
    \draw [thick, dashed] (3, 0) -- (4.5, 0);
    \draw [thick, densely dotted] (4.5, 0) -- (6, 0);
    \draw [thick, solid] (0,0) .. controls (-0.5,2) and (1.5,2) .. (3, 0);
    \draw [thick, solid] (1.5, 0) .. controls (1,-2) and (3,-2) .. (4.5, 0);

    \node[mydot, label=below:$q_0$] at (0, 0) {};
    \node[mydot, label=above:$q_1$] at (1.5, 0) {};
    \node[mydot, label=below:$q_2$] at (3, 0) {};
    \node[mydot, label=above:$q_3$] at (4.5, 0) {};
    \node[mydot, label=below:$q_4$] at (6, 0) {};
\end{tikzpicture}\quad\rightarrow\quad
	\begin{cases}
		\Gamma_1 = \left\{\swapop{5}(0, 1), \swapop{5}(2, 3)\right\},\\
		\Gamma_2 = \left\{\swapop{5}(1, 2), \swapop{5}(3, 4)\right\},\\
		\Gamma_3 = \left\{\swapop{5}(0, 2), \swapop{5}(1, 3)\right\}\,.
	\end{cases}$
    \caption{An example of edge coloring and the corresponding partitioning of the generators set.
	On the LHS we have the graph $\mathcal{M}$ with the line pattern representing the color of the edge.}
	\label{fig:edge-coloring-to-gens}
\end{figure}

We recall that the $m$ swap targets $i, j$ for each swap $\swapop{m}(i, j) \in \mathcal{P}_{\mathcal{M}}$,
and the swaps in $\mathcal{P}_{\mathcal{M}}$, correspond respectively to the vertices and the edges of the hardware connectivity graph $\mathcal{M}$.
Then the minimum number of partitions of $\mathcal{M}$, such that no two incident edges are in the same subset, is called the \textit{edge chromatic index}.
For a graph $\mathcal{G}$ we denote its chromatic index with $\chridx(\mathcal{G})$.
Therefore the optimal partitions $\Gamma_k$ correspond to the colors in the edge coloring problem.
An example for this general case is provided in Figure \ref{fig:edge-coloring-to-gens}.

Assume that the hardware graph $\mathcal{M}$ corresponds to an arbitrary topology.
Then, by applying the edge coloring procedure, we determine the optimal partitioning
\begin{align}
	\mathcal{P}_\mathcal{M} =& \bigcup_{k=1}^{\chridx(\mathcal{M})} \Gamma_k,
\end{align}
such that fixed any $k \in \intset{1}{\chridx(\mathcal{M})}$, we have that $P_a, P_b \in \Gamma_k$ implies $P_a P_b = P_b P_a$.

Given the max degree of a graph $\mathcal{G}$, denoted by $\gmaxdeg(\mathcal{G})$, a well known result by Vizing \cite{VIZING} establishes
the boundaries for its chromatic index. We report the aforementioned theorem.
\begin{theorem}[Vizing]
	\label{theorem:vizing}
	Given any finite and simple graph $\mathcal{G}$, then $\gmaxdeg(\mathcal{G}) \le \chridx(\mathcal{G}) \le \gmaxdeg(\mathcal{G}) + 1$.
\end{theorem}
In other words we need at most $\gmaxdeg(\mathcal{M}) + 1$ subsets for the partitioning of $\mathcal{P}_\mathcal{M}$.

In relation to the heavy-hex lattice topology depicted in Figure \ref{fig:heavy-hex-lattice},
we notice that graphs of such form are planar and have maximum degree 3.
Consequently, by Theorem \ref{theorem:vizing}, a hexagonal lattice connectivity requires not more than 4 partitions.

\subsection{Proofs}
\label{section:proofs}

\begin{proof}[Proof of Lemma \ref{lemma:vec-identities}]
	The first equality follows immediately from the definitions of $\vecop(\idenm{n})$ and $\vecop(A)$.
	For the second one, consider the decomposition $A=\sum_{i, j} A_{i, j} \ket{i}_n\bra{j}_n$, then
	\begin{subequations}
	\begin{align}
		\left(\idenm{n}\otimes A^\top\right) \vecop(\idenm{n}) =&
		\left(\idenm{n}\otimes \sum_{i, j} A_{i, j} \ket{j}_n\bra{i}_n\right) \sum_{k} \ket{k}_n\otimes \ket{k}_n\\
		=& \sum_{i, j} \left(\ket{i}_n\otimes  \left(A_{i, j} \ket{j}_n\right)\right) = \vecop(A)\,.
	\end{align}
	\end{subequations}
\end{proof}

\begin{proof}[Proof of Lemma \ref{lemma:vec-tp-hp}]
	We prove the equality between the left-hand side (LHS) of \eqref{eq:bellt-atpb-bell} and \eqref{eq:onest-ahpb-ones}, so
	\begin{subequations}
	\begin{align}
		\vecop(\idenm{n})^\top (A\otimes B) \vecop(\idenm{n}) =&
		\left(\sum_{i} \bra{i}\otimes \bra{i}\right) (A\otimes B) \left(\sum_{j} \ket{j}\otimes \ket{j}\right)\\
		=& \sum_{i, j} \bra{i}A\ket{j} \otimes \bra{i}B\ket{j}\\
		=& \sum_{i, j} A_{i, j} B_{i, j} = \onesv{n}^\top (A\odot B) \onesv{n}\,.
	\end{align}
	\end{subequations}
	Using Lemma \ref{lemma:vec-identities}, we prove the trace identity.
	First note that, for some $i, j \in \intset{0}{n-1}$,
	$\ket{j} \otimes \left(\bra{i}\idenm{n}\ket{j}\right)=\ket{i}$ when $i=j$ and $0$ otherwise.
	Then
	\begin{subequations}
	\begin{align}
		\traceop\left(AB^\top\right) =& \sum_{i=0}^{n-1} \bra{i} AB^\top \ket{i}\\
		=& \sum_{i, j=0}^{n-1} \bra{i} AB^\top \ket{j} \otimes \bra{i}\idenm{n}\ket{j}\\
		=& \vecop(\idenm{n})^\top (AB^\top \otimes \idenm{n}) \vecop(\idenm{n})\\
		=& \vecop(\idenm{n})^\top (A \otimes \idenm{n})(B^\top \otimes \idenm{n}) \vecop(\idenm{n})\\
		\overset{\eqref{eq:vec-identities}}{=}& \vecop(\idenm{n})^\top (A \otimes \idenm{n})(\idenm{n} \otimes B) \vecop(\idenm{n})\\
		=& \vecop(\idenm{n})^\top (A \otimes B)\vecop(\idenm{n})\,.
	\end{align}
	\end{subequations}
\end{proof}

\begin{proof}[Proof of Lemma \ref{lemma:dsm-p-tp-p}]
	Since $Q$ is doubly stochastic, then $\sum_i \lambda_i=1$ and $\lambda_i \ge 0$.
	We verify the Definition \ref{def:dsm}, so
	\begin{subequations}
	\begin{align}
		K(J_m\otimes J_m) =& \left(\sum_i \lambda_i P_i \otimes P_i\right) (J_m \otimes J_m)\\
		=& \sum_i \lambda_i (P_i J_m) \otimes (P_i J_m)\\
		=& \left(\sum_i \lambda_i\right) J_m \otimes J_m = J_m \otimes J_m\,.
	\end{align}
	\end{subequations}
	Similarly, it can be proved for the left-hand side multiplication by $J_m \otimes J_m$.
	Thus $K(J_m \otimes J_m) = (J_m \otimes J_m)K = J_m \otimes J_m$, hence the implication follows.
\end{proof}

\begin{lemma}
	\label{lemma:pdv-ell}
	The partial derivatives of the hardware cost function \eqref{eq:circcostfun:psw} w.r.t. $\timeent{\theta_s}$,
	follow the \underline{parameter shift rule} \cite{param-shift-rule}, that is
	\begin{align}
		\pdv{\mathcal{L}(\boldsymbol{\theta})}{\timeent{\theta_s}}
		=& \mathcal{L}\left(\boldsymbol{\theta} + \frac{\pi}{4}\timeent{\mathbf{e}_s}\right)
		- \mathcal{L}\left(\boldsymbol{\theta} - \frac{\pi}{4}\timeent{\mathbf{e}_s}\right),
	\end{align}
	where $\timeent{\mathbf{e}_s}$ is the standard basis vector whose index is $(t, s)$.
\end{lemma}
\begin{proof}
	First note that
	\begin{subequations}
	\begin{align}
		\pdv{\sin^2(\theta)}{\theta} =& 2 \sin\theta \cos\theta = \sin(2\theta)\\
		=& \left(\frac{1}{2} + \frac{1}{2}\sin(2\theta)\right) - \left(\frac{1}{2} - \frac{1}{2}\sin(2\theta)\right)\\
		=& \left(\frac{1}{2} - \frac{1}{2}\cos(2\theta + \frac{\pi}{2})\right)
		- \left(\frac{1}{2} - \frac{1}{2}\sin(2\theta - \frac{\pi}{2})\right)\\
		\label{eq:sin-twotheta-as-diff-sin}
		=& \sin^2\left(\theta + \frac{\pi}{4}\right) - \sin^2\left(\theta - \frac{\pi}{4}\right)\,.
	\end{align}
	\end{subequations}
	Consider the matrix-valued function $\mathrm{SSWAP}_m((\cdot); i, j): \mathbb{R} \to \mathcal{M}_m$, then
	fixed some $\mathbf{v}, \mathbf{w} \in \mathbb{R}^m$, we have
	\begin{subequations}
	\begin{align}
		\label{eq:pdv-ell-start}
		\pdv{\left(\mathbf{v}^\top \mathrm{SSWAP}_m(i, j, \theta) \mathbf{w}\right)}{\theta}
		=& \pdv{\sin^2(\theta)}{\theta} \left(\mathbf{v}^\top\swapop{m}(i, j)\mathbf{w} - \mathbf{v}^\top \mathbf{w}\right)\\
		=& \mathbf{v}^\top\mathrm{SSWAP}_m\left(i, j, \theta + \frac{\pi}{4}\right)\mathbf{w}
		- \mathbf{v}^\top\mathrm{SSWAP}_m\left(i, j, \theta - \frac{\pi}{4}\right)\mathbf{w},
	\end{align}
	\end{subequations}
	and similarly for $\mathrm{PSSWAP}$.
	But the LHS of \eqref{eq:pdv-ell-start} corresponds to the form of $\pdv{\mathcal{L}_t(\vectheta{})}{\timeent{\theta_s}}$.
	Also, since the variables $\timeent{\theta_s}$ appear not more than once in each term of the cost in
	\eqref{eq:circcostfun:psw} then by linearity the result follows.
\end{proof}

\begin{lemma}
	\label{lemma:dsm-preserve-l1}
	Let $Q$ be an $m\times m$ DSM, then the $\ell_1$-norm of vectors from the non-negative orthant
	$\mathbb{R}^m_+$ is preserved under the action of $Q$. That is
	\begin{align}
		\mathbf{v} \in \mathbb{R}^m_+ \implies& \|Q\mathbf{v}\|_1=\|\mathbf{v}\|_1,
	\end{align}
	for all $m\times m$ DSMs $Q$.
\end{lemma}
\begin{proof}
	Since $\mathbf{v} \in \mathbb{R}^m_+$, so each component $v_i$ is non-negative, then the $\ell_1$-norm can be written as
	$\|\mathbf{v}\|_1 = \onesv{m}^\top \mathbf{v}$.
	Note that for any $m\times m$ permutation matrix $P$, we have $P\onesv{m}=\onesv{m}$.
	Let $Q=\sum_j \lambda_j P_j$ be any $m\times m$ DSM, where $\lambda_j$ and $P_j$ follow Definition \ref{def:dsm}.
	Hence
	\begin{subequations}
	\begin{align}
		\|Q\mathbf{v}\|_1 =& \left\|\sum_j \lambda_j P_j \mathbf{v}\right\|_1
		= \sum_j \lambda_j \onesv{m}^\top P_j \mathbf{v}\\
		=& \onesv{m}^\top \mathbf{v} \sum_j \lambda_j = \|\mathbf{v}\|_1,
	\end{align}
	\end{subequations}
	consequently the claim is proved.
\end{proof}

\begin{proof}[Proof of Proposition \ref{lemma:ell-critical-points}]
	From Lemma \ref{lemma:pdv-ell} and the structure of the function $\mathcal{L}$,
	it follows that the partial derivative w.r.t. $\timeent{\theta_s}$ takes the form
	\begin{subequations}
	\begin{align}
		\label{eq:pdv-zeros-1}
		\pdv{\mathcal{L}(\boldsymbol{\theta})}{\timeent{\theta_s}}
		=& \sum_{\tau=t}^{T-1} \mathbf{v}_\tau^\top
		\left(\idenm{m}^{\otimes 2} + \sin^2\left(\timeent{\theta_s} + \frac{\pi}{4}\right) F\right)
		\mathbf{w}_\tau\\
		\nonumber
		-& \sum_{\tau=t}^{T-1} \mathbf{v}_\tau^\top
		\left(\idenm{m}^{\otimes 2} + \sin^2\left(\timeent{\theta_s} - \frac{\pi}{4}\right) F\right)
		\mathbf{w}_\tau,
	\end{align}
	with
	\begin{align}
		F =& \left(\timeent{S_s}\right)^{\otimes 2} - \idenm{m}^{\otimes 2},
	\end{align}
	\end{subequations}
	where $\mathbf{v}_\tau, \mathbf{w}_\tau$ are some fixed vectors and $\timeent{S_s}$ is the swap corresponding to the
	parameter $\timeent{\theta_s}$. We note that the vectors $\mathbf{v}_\tau, \mathbf{w}_\tau$ may depend on the elements of
	$\boldsymbol{\theta}$ excluding the selected $\timeent{\theta_s}$.
	Equating \eqref{eq:pdv-zeros-1} with zero we obtain
	\begin{align}
		\label{eq:pdv-zeros-2}
		\sin^2\left(\timeent{\theta_s} + \frac{\pi}{4}\right)
		\sum_\tau \left(\mathbf{v}_\tau^\top F \mathbf{w}_\tau\right)
		=& \sin^2\left(\timeent{\theta_s} - \frac{\pi}{4}\right)
		\sum_\tau \left(\mathbf{v}_\tau^\top F \mathbf{w}_\tau\right),
	\end{align}
	then by \eqref{eq:sin-twotheta-as-diff-sin}, we immediately see that the latter holds when
	$\sin\left(2 \timeent{\theta_s}\right)=0$, that is $\timeent{\theta_s} \in \Omega$.
	Consequently, points \textit{1.} and \textit{2.} follow.

	For point \textit{3.}, it is sufficient to show that $\mathbf{v}_\tau^\top F \mathbf{w}_\tau$ can be zero, then
	the value of $\timeent{\theta_s}$ is uninfluential. Therefore, the partial derivative w.r.t. the same variable is zero,
	independently from the value of $\timeent{\theta_s}$.

	Vectors $\mathbf{v}_\tau, \mathbf{w}_\tau$ are the result of some DSM applied to the non-zero vectors,
	with non-negative entries, $\vecop\left(\timeent{G}\right)$ and $\vecop\left(M_c\right)$.
	By Lemma \ref{lemma:dsm-preserve-l1} the DSM action on those vectors preserves the $\ell_1$-norm,
	then $\mathbf{v}_\tau\ne \mathbf{0}, \mathbf{w}_\tau\ne \mathbf{0}$.
	Thus $\mathbf{v}_\tau^\top F \mathbf{w}_\tau=0$ only as a result of the action of $F$. We show that there exist basis vectors $\ket{a}_m, \ket{b}_m$
	such that the latter equality holds. Assume $\timeent{S_s}=\swapop{m}(i, j)$ for some $i\ne j$ such that $i, j$ are distinct from $a, b$, then
	\begin{subequations}
	\begin{align}
		\bra{a}_m^{\otimes 2} F \ket{b}_m^{\otimes 2}
		=& \bra{a}_m^{\otimes 2}\swapop{m}(i, j)^{\otimes 2} \ket{b}_m^{\otimes 2} -
		\bra{a}_m^{\otimes 2} \ket{b}_m^{\otimes 2} =0,
	\end{align}
	\end{subequations}
	since $\swapop{m}(i, j)$ fixes $\ket{a}_m, \ket{b}_m$.
	Hence there exist $\mathbf{v}_\tau, \mathbf{w}_\tau$ such that $\mathbf{v}_\tau^\top F \mathbf{w}_\tau$ vanishes, so point \textit{3.} is proved.
\end{proof}

\begin{lemma}
	\label{lemma:min-card}
	Consider the $C^1$ function $g:\mathbb{R}^n \to \mathbb{R}_+$, such that
	$g(\mathbf{x}^*)=0 \implies \mathbf{x}^* \in \mathbb{Z}^n$
	and $g(\mathbf{x} + 2\mathbf{e}_i)=g(\mathbf{x})$ for all $i=0\isep n-1$, with $\{\mathbf{e}_i\}$ the canonical basis for $\mathbb{R}^n$.
	In other words the zeros of $g$ occur at points having integer entries, additionally the function is entry-wise periodic with \textit{fundamental period} 2.
	Then the following optimization problems are equivalent
	\begin{equation}
		\label{lemmal2card}
		\begin{cases}
			\underset{\mathbf{w} \in \mathbb{R}^n}{\mathrm{min}} & \|\mathbf{w}\|^2_2,\\
			\text{s.t.}\quad & g(\mathbf{w}) = 0,
		\end{cases}
		\quad \cong \quad
		\begin{cases}
			\underset{\mathbf{w} \in \mathbb{R}^n}{\mathrm{min}} & \mathbf{card}(\mathbf{w}),\\
			\text{s.t.}\quad & g(\mathbf{w}) = 0,
		\end{cases}
	\end{equation}
	where $\mathbf{card}(\cdot)$ denotes the cardinality of the argument.
\end{lemma}
\begin{proof}
	A consequence of the periodicity is that for any $\mathbf{x} \in \mathbb{R}^n$ and
	$\mathbf{y}=\left(2\sum_i k_i \mathbf{e}_i\right) \in 2\mathbb{Z}^n$, with $k_i \in \mathbb{Z}$, we have $g(\mathbf{x}+\mathbf{y})=g(\mathbf{x})$.

	Let $\mathbf{x}^* \in \mathbb{Z}^n$ fixed, and consider the \textit{projection}\footnote{
		We highlight a technicality regarding the projection $\projop{2\mathbb{Z}}$.
		Since the set $2\mathbb{Z}^n$ is non-empty closed but non-convex, then we cannot assure that the projection is
		a singleton \cite[first projection theorem]{BeckBook},
		consequently we assume $\projop{2\mathbb{Z}}$ to be a set-valued operator, with the non-emptiness resulting from the
		set $2\mathbb{Z}^n$ being non-empty and close.
	} $\projop{2\mathbb{Z}}$
	of $\mathbf{x}^*$ onto the set $2\mathbb{Z}$, so let
	\begin{align}
		\label{l2ystarargmin}
		\mathbf{y}^* \in \projop{2\mathbb{Z}}(\mathbf{x}^*)
		= \underset{\mathbf{y} \in 2\mathbb{Z}^n}{\mathrm{argmin}}\, \left\{\|\mathbf{x}^* - \mathbf{y}\|^2_2\right\},
	\end{align}
	then $\mathbf{x}^* - \mathbf{y}^* \in \{-1, 0, 1\}^n$, so
	\begin{subequations}
	\begin{align}
		\label{l2xstartpystar}
		\|\mathbf{x}^* - \mathbf{y}^*\|^2_2
		=& \sum_i (x^*_i - y^*_i)^2
		= \sum_i \left|x^*_i - y^*_i\right|\\
		=& \|\mathbf{x}^* - \mathbf{y}^*\|_1\\
		=& \mathbf{card}(\mathbf{x}^* - \mathbf{y}^*)
	\end{align}
	\end{subequations}
	Now, consider the first optimization problem in \eqref{lemmal2card} and perform the substitution $\mathbf{w}=\mathbf{x}-\mathbf{y}$,
	to obtain
	\begin{align}
		\underset{\mathbf{x} \in \mathbb{R}^n, \mathbf{y} \in 2\mathbb{Z}^n}{\mathrm{min}} &
		\|\mathbf{x}-\mathbf{y}\|^2_2,\\
		\nonumber
		\text{s.t.}\quad & g(\mathbf{x} - \mathbf{y})=g(\mathbf{x}) = 0\,.
	\end{align}
	Let $G=\{\mathbf{x} \in \mathbb{R}^n| g(\mathbf{x})=0\}$ be the set of feasible points,
	then by assumption, $\mathbf{x} \in G \implies \mathbf{x} \in \mathbb{Z}^n$, that is $G \subset \mathbb{Z}^n$.
	Since the constraint is independent from the variable $\mathbf{y}$,
	we rewrite the optimization problem considering the feasible set, so
	\begin{align}
		\underset{\mathbf{x} \in G, \mathbf{y} \in 2\mathbb{Z}^n}{\mathrm{min}}\quad
		\|\mathbf{x}-\mathbf{y}\|^2_2,
		& \quad\cong\quad
		\underset{\mathbf{y} \in 2\mathbb{Z}^n}{\mathrm{min}}\quad
		\underset{\mathbf{x} \in G}{\mathrm{min}}\quad
		\|\mathbf{x}-\mathbf{y}\|^2_2,
	\end{align}
	then by \eqref{l2ystarargmin} and \eqref{l2xstartpystar}, the optimal $\mathbf{y}^*$ is the one such that
	$\mathbf{x}^* - \mathbf{y}^* \in \{-1, 0, 1\}^n$ and the cardinality of $\mathbf{x}^* - \mathbf{y}^*$ is minimized.
	Hence the equivalence in \eqref{lemmal2card} is established.
\end{proof}

\begin{proof}[Proof of Proposition \ref{thm:equiv-optim-prob}]
	We sketch a proof based on Lemma \ref{lemma:min-card}.
	The aforementioned lemma can be adapted to the periodicity of the hardware cost function $\mathcal{L}$.
	It follows from \eqref{eq:psswapdef} that the period of $\mathcal{L}$ is $\pi$.
	Define the function
	\begin{align}
		g(\mathbf{x})=&\mathcal{L}\left(\frac{\pi \mathbf{x}}{2}\right),
	\end{align}
	which has period 2.
	Also the cardinality function is invariant to a non-zero vector scaling, that is $\cardf(\mathbf{x})=\cardf(\alpha \mathbf{x})$
	for all $\alpha \in \mathbb{R}\setminus\{0\}$.
	Then by applying Lemma \ref{lemma:min-card}, the claim follows.
\end{proof}

\begin{proof}[Proof of Lemma \ref{lemma:commuting-perm-mul-as-sum}]
	Under the assumptions that $P_a^2=P_b^2=\idenm{m}$ and $P_aP_b=P_bP_a$ we obtain that
	$\left(P_aP_b\right)^2=P_aP_bP_aP_b=P_aP_bP_bP_a=\idenm{m}$, then we need that
	\begin{align}
		(P_a + P_b - \idenm{m})^2 =& \idenm{m} + 2\left[P_aP_b - (P_a + P_b - \idenm{m})\right] = \idenm{m},
	\end{align}
	which is true if and only if $P_aP_b = (P_a + P_b - \idenm{m})$.
	The latter being the claim proves the lemma.
\end{proof}

\begin{proof}[Proof of Proposition \ref{thm:dsm-comp-gen-commuting-perm-as-sum}]
	First we note that, for $m\times m$ permutation matrices $P_a, P_b$, fulfilling the conditions of Lemma \ref{lemma:commuting-perm-mul-as-sum},
	then
	\begin{align}
		\label{eq:comm-perm-shift-iden-comp-vanish}
		(P_a - \idenm{m})(P_b - \idenm{m}) =& P_a P_b - P_a - P_b + \idenm{m}
		\overset{\eqref{eq:commuting-perm-mul-as-sum}}{=} 0\,.
	\end{align}
	In other words, given any pair of distinct, commuting and involutory permutations (acting on the same vector space),
	the product of their shift by the identity vanishes.

	We proceed by induction. Equation \eqref{eq:dsm-comp-gen-commuting-perm-as-sum} is clearly true for $n=1$,
	furthermore, assume it is true for an arbitrary $n\ge 1$,
	then consider the case $n+1$ by multiplying both sides of equation \eqref{eq:dsm-comp-gen-commuting-perm-as-sum}
	by $\idenm{m} + \sin^2(\theta_{n+1}) \left(P_{n+1} - \idenm{m}\right)$ to get
	\begin{subequations}
	\begin{align}
		\prod_{t=1}^{n+1}
		\left(\idenm{m} + \sin^2(\theta_t) \left(P_t - \idenm{m}\right)\right)
		=& \idenm{m} + \sum_{t=1}^{n+1} \sin^2(\theta_t)\left(P_t - \idenm{m}\right)\\
		& + \sum_{t=1}^{n} \sin^2(\theta_t) \sin^2(\theta_{n+1})
		\underbrace{\left(P_t - \idenm{m}\right)\left(P_{n+1} - \idenm{m}\right)}_{\text{$0$ by \eqref{eq:comm-perm-shift-iden-comp-vanish}}}\\
		=& \idenm{m} + \sum_{t=1}^{n+1} \sin^2(\theta_t)\left(P_t - \idenm{m}\right),
	\end{align}
	\end{subequations}
	hence the claim is proved.
\end{proof}


\end{document}